\documentclass[conference]{IEEEtran}

\ifCLASSINFOpdf
\else
\fi

\newcommand{\mno}{mobile network operator }
\newcommand{\mnos}{mobile network operators }

\newcommand{\Mnos}{Mobile network operators }
\newcommand{\mnonospace}{mobile network operator}
\newcommand{\mnosnospace}{mobile network operators}

\usepackage{circuitikz}
\usepackage{tikz}
\usepackage{amsmath}
\usepackage{hyperref}
\usepackage{csquotes}
\usepackage{subcaption}
\usepackage{multirow}
\usepackage{csquotes}
\usepackage{comment}
\usepackage{paralist}
\usepackage{url}

\usepackage[shortlabels]{enumitem}

\newif\ifstatus
\statustrue
\begin{document}

\title{An Overview of 7726 User Reports:\\
Uncovering SMS Scams and Scammer Strategies}

% \author{Anonymous Author(s)}

\author{\IEEEauthorblockN{Sharad Agarwal}
	\IEEEauthorblockA{University College London (UCL)\\
		sharad.agarwal@ucl.ac.uk}
	\and
	\IEEEauthorblockN{Guillermo Suarez-Tangil}
	\IEEEauthorblockA{IMDEA Networks\\
		guillermo.suarez-tangil@networks.imdea.org}
	\and
	\IEEEauthorblockN{Marie Vasek}
	\IEEEauthorblockA{University College London (UCL)\\
		m.vasek@ucl.ac.uk}}

% \IEEEoverridecommandlockouts
% \makeatletter\def\@IEEEpubidpullup{6.5\baselineskip}\makeatother
% % \IEEEpubid{\parbox{\columnwidth}{
% % 		Network and Distributed System Security (NDSS) Symposium 2026\\
% % 		23-27 February 2026, San Diego, CA, USA\\
% % 		ISBN 979-8-9894372-8-3\\
% % 		https://dx.doi.org/10.14722/ndss.2026.240987\\
% % 		www.ndss-symposium.org
% % }
% \hspace{\columnsep}\makebox[\columnwidth]{}}

% make the title area
\maketitle

\begin{abstract}
\Mnos implement firewalls to stop illicit messages, but scammers find ways to evade detection. Previous work has looked into SMS texts that are blocked by these firewalls. However, there is little insight into SMS texts that bypass them and reach users. To this end, we collaborate with a major %UK 
\mno to receive 1.35$m$ user reports submitted
%to the SMS reporting service, 7726,
over four months. 
We find 89.16\% of user reports comprise text messages, followed by reports of suspicious calls and URLs. 
Using our methodological framework, we identify 35.12\% of the unique text messages reported by users as spam, while 40.27\% are scam text messages.
This is the first paper that investigates SMS reports submitted by users and differentiates between spam and scams.
Our paper classifies the identified scam text messages into 12 scam types, of which the most popular is `wrong number' scams. 
 %, and discusses the
We explore the various infrastructure services that scammers abuse to conduct SMS scams, including \mnos and hosting infrastructure, and 
%We also
analyze the text of the scam messages to understand how scammers lure victims into providing them with their personal or financial details.
%The mobile network analysis indicates that the scammers mainly abuse the four major UK \mnos to send scam text messages, with a median lifetime of 48 days. We also find that scammers abuse URL shorteners and proxy services like Cloudflare to evade detection and present the details of the hosting infrastructure they abuse to conduct scams. 
%The text analysis indicates that scammers use the kindness and distraction principle for conversational scams like the Wrong Number scam and the time/urgency principle for all scams. 
%The Gunning Fog Index shows that the readability for Wrong Number and Hi Mum/Dad scams is much lower than other scam types. 

\end{abstract}

% For peer review papers, you can put extra information on the cover
% page as needed:
% \ifCLASSOPTIONpeerreview
% \begin{center} \bfseries EDICS Category: 3-BBND \end{center}
% \fi
%
% For peerreview papers, this IEEEtran command inserts a page break and
% creates the second title. It will be ignored for other modes.
\IEEEpeerreviewmaketitle

\section{Introduction}
\label{sec:intro}

There has been a recent surge in SMS scams worldwide~\cite{trendmicro_scam, techradar_smsscam}, with over 300$k$ fraudulent texts sent everyday~\cite{proofpoint_mwc}. %The Federal Trade Commission (FTC) reported 
Unlike phishing, where data is easily accessible through aggregators such as OpenPhish~\cite{openphish}, Phishtank~\cite{phishtank} and APWG eCX~\cite{apwg_ecx}, studying SMS scams is a lot harder due to unavailability of updated public data. Blocking scam texts is not universally implemented by \mnos and the limited metadata available in SMS --- sender ID and timestamp, makes detection challenging. 
In 2024, users in the US lost \$470$m$ to text scams~\cite{ftc_smsscams}, %with more than %59$k$ complaints resulting in over 
including \$129$k$ attributed to just toll-related smishing campaigns~\cite{fbi_ic3:24}. 
%The FBI's IC3 annual report highlights more than 59$k$ complaints resulting in over \$129$k$ losses to just toll-related smishing texts~\cite{fbi_ic3:24}.
%
%SMS scams have surged in the UK, impacting the country's economy with more than £74.78$m$ lost to authorized push payment (APP) fraud enabled by telecommunications in the first half of 2024~\cite{ukfinance_2024}.
%
In the same period, the UK reported an 8\% increase in SMS scams~\cite{gasa_uk}, %costing individuals over 
contributing to more than £162$m$ in losses from telecommunications-enabled authorized push payment (APP) fraud~\cite{ukfinance_2025}.
%With an 8\% rise in SMS scams~\cite{gasa_uk}, the UK's economy was impacted with more than £162$m$ lost 
%to telecommunication enabled authorized push payment (APP) fraud~\cite{ukfinance_2025}. %enabled by telecommunications in 
%the first half of 
%2024~\cite{ukfinance_2025}.
%The Global Anti Scam Alliance (GASA) reported an 8\% rise in SMS scams in the UK~\cite{gasa_uk}. 
%impacting the country's economy \marie{either motivate the impact to the economy or omit the clause}. UK Finance reports a loss of £571.7 million in the first half of 2024, out of which £213.7 million was due to advanced push payment (APP) fraud. 16\% of APP is initiated through telecommunications networks, leading to a loss of over 35\% losses \marie{this makes no sense. rewrite.} (i.e., £74.8 million)~\cite{ukfinance_2024}.
%According to the UK Home Office,
While Australia recorded over \$14$m$ in losses due to smishing in 2024~\cite{scamwatch_aus}, fraudsters stole over \$4.2$m$ in just three months from users in New Zealand~\cite{nz_loss}.
Notably, most scams %in the UK 
are delivered via text messages and phone calls~\cite{gasa_uk}.
%This rises to a significant amount of criminal activity: 40\% of all crime in England and Wales is fraud~\cite{uk_homeoffice}.
%
%Phishing is ranked the most reported scam per the latest FBI IC3 report~\cite{fbi_ic3}. \marie{this last sentence doestn fit. consider omitting.}

%{\color{red} 
Despite the substantial financial losses caused by SMS scams, 
most government (or commercial) reports fail to analyze and publish detailed reports on SMS scams. 
For example, the FBI IC3 annual report groups smishing, vishing, and phishing into one, without differentiation~\cite{fbi_ic3}. 
The lack of insights into the SMS scam ecosystem can be attributed to the difficulties organizations face in data sharing. %}

Government telecom regulators in %various countries such as 
the US, Canada, UK, New Zealand, and Australia work with \mnos to combat the increasing amount of telecom-related scams. To this end, \mnos %in the UK
have implemented SMS firewall filters and detection systems~\cite{canada:mno_filter, spamshield_mavenir, nz:mno_filter, aus:mno_filter} to stop SMS scams, including smishing  
%Smishing is the act of deceiving an individual by luring them into clicking on a malicious URL to gain financial or sensitive details.
(SMS phishing). %text message).
Similar to phishing, this has become a cat-and-mouse game where scammers create new URLs~\cite{agarwal25:weis}, and \mnosnospace, along with threat intelligence organizations, detect and block known URLs and text messages containing them.
Despite the \mnonospace's filters blocking malicious messages, scammers evade detection by changing sender IDs, message text, and URLs.
Prior work has investigated the texts blocked by these filters~\cite{agarwal_imc24} and we lack insights into scam texts that evade detection and reach users.

% Scams are more commonly attempted via text messages, with 7 in 10 people receiving a suspicious text~\cite{ofcom_21}. %In the UK, 
% Alarmingly, 1 in 10 people have fallen victim to such scams~\cite{cifas}. %in the UK%, as per a study conducted by GASA and CIFAS
% %Ofcom reports that 
% %Related work found 
% Prior studies found that 17\% of participants fell for a smishing attack~\cite{lutforsmishing_interview} and that users focus on the SMS content rather than the sender ID to identify smishing~\cite{smishinguserstudy_soups24}.

As \mnos continue to block scams, threat actors find new tactics to evade detection and deceive users. 
One such technique includes sending a text message and asking users to call or text back. %on a phone number. 
Others found that prompting users to reply and click on a URL increases the odds of a %user responding 
response to a smishing text~\cite{lutforsmishing_interview}. 
While 7 in 10 people receive a suspicious text~\cite{ofcom_21}, 1 in 10 fall victim~\cite{cifas}. 
Controlled studies show at least 17\% of participants fall for a smishing attack~\cite{lutforsmishing_interview, timko2025understanding} and users focus on the SMS content rather than the sender ID to identify smishing~\cite{smishinguserstudy_soups24}. 
To devise effective countermeasures, it is essential to understand the various successful types of scams and how scammers deceive their victims. 

\Mnos %in certain countries, like the UK,
in %most English-speaking countries
the US, Canada, UK, and New Zealand run a special reporting service called 7726 (`SPAM' on a mobile keypad), which allows users to report suspicious calls and SMS messages~\cite{ofcom_7726, att_7726, nz_7726, ca_7726}. 
%This allows users to report %suspicious text messages and calls that
Reported messages have evaded the \mnonospace's filters, and this service helps \mnos update their detection rules with new threats. %to detect new types of scams. 
%For user convenience,
\Mnos collaborate with Google and Apple to integrate one-click reporting into the 7726 data feed.
As users confuse legitimate and illegitimate text messages~\cite{timko2023more}, the reports submitted to 7726 are undifferentiated and contain a mix of spam, scams, and legitimate texts.
%need investigation
%to differentiate between spam, scam, and legitimate texts. %\marie{idk whst distinction youre making here between real and fake.}

\vspace{.1cm}\noindent\textbf{Research Gap.}
\Mnos have %in the UK 
implemented an SMS detection system to reduce the amount of smishing received by end users. %However, criminals evade the detection of these firewalls, in which case, the victim receives the SMS. 
While previous research has examined these blocked SMS texts to study an individual scam~\cite{agarwal_usenix25} or identify scam types~\cite{agarwal_imc24}, 
%Agarwal et al. collaborate with a UK \mno to access blocked `hi mum and dad' SMS text and investigate the conversational scam by interacting with scammers~\cite{agarwal_usenix25}.
%In another paper, they access two months of all blocked SMS texts by a UK \mno and identify seven categories of SMS scams and spam~\cite{agarwal_imc24}. 
%\marie{i dont understand why this isnt in the section on other related works. its too much of a new leap. you need to set it up more earlier.}
%While the previous research has provided insights into blocked SMS messages, 
it is essential to understand the text messages that bypass the filters and get delivered to the users. The data collected by previous studies through online forums~\cite{tang:ccs22}, public online SMS gateways~\cite{nahapetyan2024sms} or crowdsourcing~\cite{smishtank_timko24} does not differentiate between scams and spam. This distinction is vital to build better mechanisms to thwart monetarily damaging cybercrime, as scams pose a significantly greater threat --- they manipulate trust to inflict harm, making their detection and prevention a priority. 
%and either has old URLs that have not been resolved in real-time or is a small dataset, making it a challenge to understand how scammers lure victims into different scams.

\vspace{.1cm}\noindent\textbf{Contribution.}
To fill this research gap, we collaborate with a major \mno that provides four months of user reports weekly. Our collaborator integrated Google's one-click SMS reports during the data collection period. We investigate 1.35 million user reports, identify scam and spam text messages, and provide insights into the identified scams.
%We also find rich communication service (RCS) smishing campaigns in these reports, pointing towards a shift of scam messages being delivered via RCS instead of SMS to evade detection by \mnonospace's firewalls.

With this data, we ask the following research questions: 

\begin{enumerate}[\bf RQ1]\label{rq}
    \setlength\itemsep{0em}
    \item\label{rq1} What kind of reports do users submit to 7726?
    \item\label{rq2} What scams exist in the reported SMS messages?
    \item\label{rq3} Which \mnos are abused, and how long scammers use these mobile numbers?
    \item \label{rq4} What infrastructure is abused by scammers to run smishing campaigns?
    \item\label{rq5} How do scammers try to lure victims?
\end{enumerate}

%The paper makes the following contributions:
In answering these RQs, we contribute the following:

\begin{itemize}
    \setlength\itemsep{0em}
    \item We investigate 1,349,039 user reports over four months in the UK and present the distribution in \S\ref{subsubsec:reports_over_time}. %of SMS texts and calls in the UK over four months. 
    %\S\ref{subsec:characterization_user_reports} presents the categorization of SMS user reports and the scams we identify. %their breakdown into suspicious calls, URLs, and text messages, further breaking it down into suspicious text messages, spam text messages, and text messages with OTPs. 
    
    %\item \S\ref{subsec:Suspicious Text Message Classification} categorizes the suspicious text messages into spam and scam. It further identifies scam types, including the ones not previously seen%

    \item We categorize SMS user reports into spam and scams, further breaking down scams into deeper subsets (\S\ref{subsec:characterization_user_reports}).

    \item We detail the infrastructure that scammers abuse to conduct SMS scams, including \mnos  (\S\ref{subsec:senderid}) and registrars (\S\ref{subsec:domain_analysis}).

    \item We investigate how scammers craft text messages, including lures that scammers use to deceive victims (\S\ref{subsec:text_analysis}).
\end{itemize}

\section{Background}
\label{sec:background}

With online messaging communication channels, there has been an overall decline in the total number of text messages sent/received. %in the UK. 
However, there has been an uptick in SMS texts for essential services such as order updates, bank transactions, and medical services. As of 2024, 97\% of adults in the UK own or have access to mobile telephone~\cite{ofcom_mobileaccess}, with 89.6$m$ active mobile subscriptions at the end of Q2 2024~\cite{ofcom_activesubscription}.

As the legitimate use of SMS increases, businesses have started using this communication channel by sending unsolicited marketing SMS texts, aka spam. %\guillermo{Revisit the use of misusing/illegitimate. With GPPR, marketing is not illegitimate if there is consent. In the US, SPAM is regulated in the CAN-SPAM Act, which makes it legal as long as you honor Unsubscriptions, etc. }
At the same time, scammers abuse SMS to deceive victims into clicking on a URL or interacting with them by impersonating brands/organizations/individuals to steal users' personal or financial details, aka scams.
While \textit{spam} texts hinder the availability of SMS texts, \textit{scam} texts intentionally cause victims financial and/or psychological harm. 
Hence, it is essential to differentiate between spam and scam texts so stakeholders, such as \mnos and governmental regulators, can take appropriate actions.
Similarly, research communities wielding this distinction will be able to develop more efficient detection models and propose more effective countermeasures to stop both spam and scams.

Recent studies on suspicious SMS messages fail to differentiate between these two types of text messages~\cite{nahapetyan2024sms,tang:ccs22,smishtank_timko24,srinivasan2016understanding}. The messages collected in previous works consist of old URLs that cannot be resolved now, and the available datasets do not include the redirected URL. Using previously collected data to differentiate spam from scams makes it challenging; \S\ref{sec:relatedwork} further discusses related academic work. 
%It will help the research community develop more efficient detection models and propose effective countermeasures to stop spam and scams. 
%\guillermo{Agreed, but you should, more importantly, explain why it is essential to the research community.}

%\guillermo{MNO firewall discussions is becoming increasingly important to the findings of our paper. We should use a more modern term to refer to the fraud solution they deploy instead of firewall. I think XDR could be a more appropriate term, or at least more closed to what industry uses nowadays for gateways that use AI and whatnot. We should also introduce the idea that they use these systems more formally. This reference down there looks very incidental. Something like this may fly:}
In response to the uptake of fraud in text messages, \mnos have implemented advanced fraud prevention systems powered by real-time AI and machine learning technologies. These systems leverage message fingerprinting algorithms to detect and mitigate messaging fraud scenarios automatically. Throughout this paper, we use the term Extended Detection and Response (XDR) system to refer to these fraud prevention systems due to its widespread use in industry. The ongoing arms race between fraudsters and \mnos significantly complicates the detection of scam messages at scale, posing a challenge even to advanced automated systems like XDR. 

In addition to their efforts to evade detection by XDR systems, scammers also carefully craft messages designed to deceive human recipients. 
They employ various lures to manipulate victims into falling for different types of scams, distinct from businesses' tactics to attract users to spam.
Understanding these lures requires access to the text being sent. This allows us to inform educational resources to inform potential victims as well as interveners like law enforcement.
%Understanding these lures requires scam text messages so educational resources and programs can effectively  potential victims.

%However, 
%Although distinguishing between spam and scam is crucial,

There exist six known SMS scam types %in the UK 
-- (1) Wrong Number, (2) Hi Mum/Dad, (3) Delivery, (4) Banking, (5) Telecom, and (6) Government~\cite{agarwal_imc24}. 
Based on these scam types,
\Mnos block text messages which their XDRs can identify. 
However, many messages (which contain unknown scams) successfully evade
%\mnonospace's
the XDRs' filters. %, which contain many more .
Our research focuses on the SMS text messages delivered to final users, evading the \mnonospace's detection capabilities.
We provide a methodological framework that systematically groups user reports, differentiates between spam and scam, and understands the lures scammers use to deceive victims into different scams.

%\guillermo{UPDATE firewall - XDR}

\section{Methodology}
\label{sec:methods}

%This section explains how we receive the 7726 user reports from our partner, a major UK \mnonospace.
%We receive 7726 user spam reports from our partner, a major UK \mnonospace.

\begin{figure}
    \centering
    \includegraphics[width=0.9\linewidth, trim=0 105 340 0, clip]{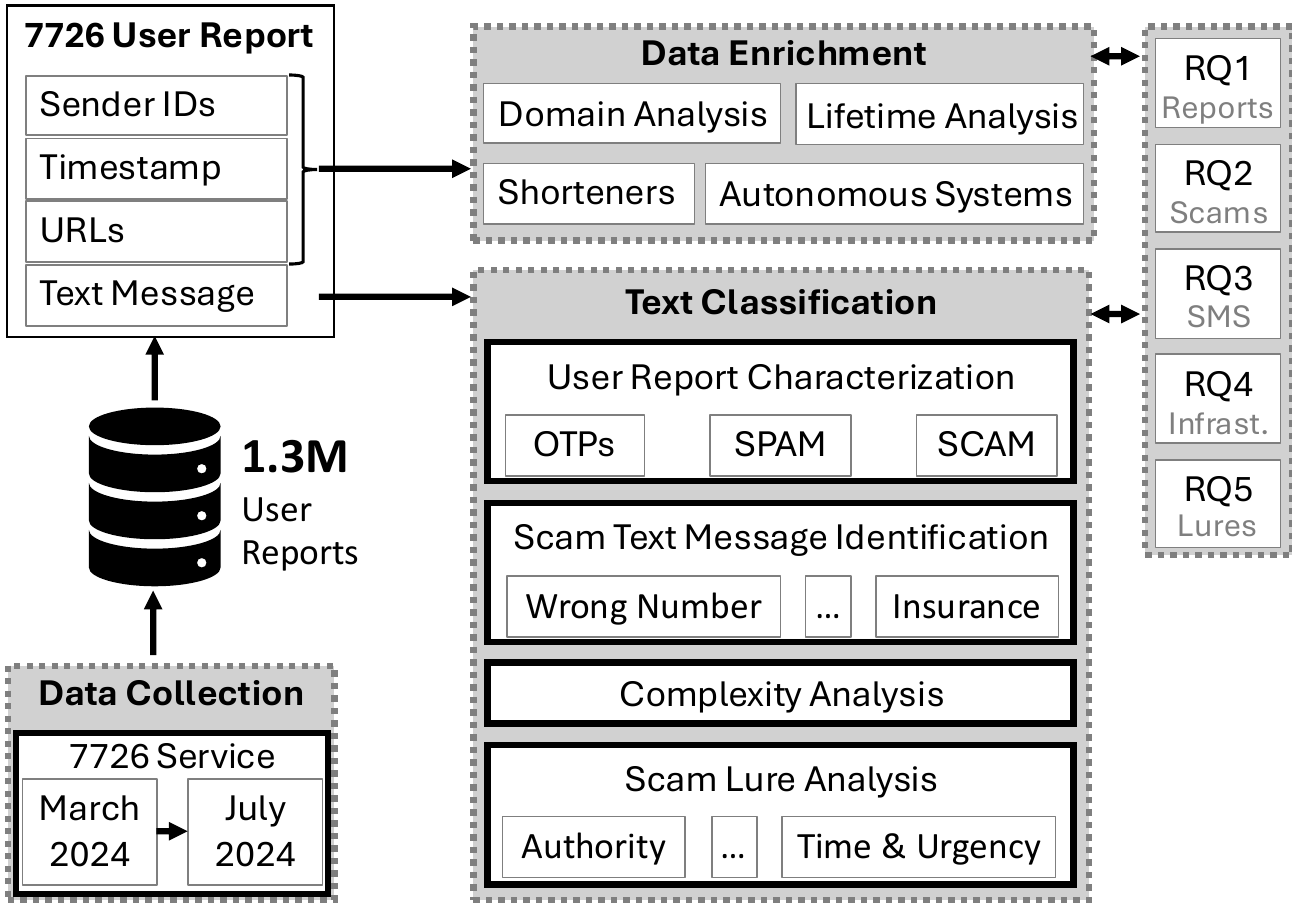}
    \caption{Overview of our processing pipeline to characterize 7726 user reports and identify SMS scams.}
    \label{fig:pipeline}
\end{figure}

\Mnos in various countries %including the United Kingdom 
run a special user reporting service --- 7726, where users can report suspicious SMS texts either by forwarding the text message or through the one-click reporting system enabled by Apple and Google in their messaging platforms.\footnote{During our data collection period, Apple one-click was not enabled by our data partner.} 
If forwarding via 7726, users are asked to share the sender ID while reporting the text message. 
The one-click solution automatically shares the sender ID with the text message.
In this section, we describe how we obtain, enrich, and examine this data to answer our research questions (\ref{rq1}-\ref{rq5}). %and the methods we use to examine user reports.
Figure~\ref{fig:pipeline} shows an overview of our processing pipeline in a nutshell. 

%\guillermo{Did we have a general figure summarizing the methodology (data enrichment, analysis, etc)? This would be a good place to place it.}

% \begin{figure}[!ht]
% \centering
% \resizebox{\linewidth}{!}{%
% \input{plots/flowchart}
% }
% \caption{Overview of our methodology to identify spam, scam, and unknown text messages from user text message reports.}
% \label{fig:sus_text_message_distribution}
% \end{figure}

\subsection{Data Collection}
\label{subsec:data collection}

%We collaborate with a major UK \mno that provides us with user reports between March 7, 2024, and July 11, 2024, every week. 
We collect weekly 7726 reports between March 7, 2024, and July 11, 2024, from a major %UK 
\mnonospace. 
We received 1,349,039 user reports consisting of SMS, URLs, and calls.  
Since our paper focuses on understanding the SMS ecosystem, we exclusively retain SMSs. 
In particular, we filter out 129,467 %(10\%)
reports flagging suspicious calls and 16,713 %(1.2\%)
reports flagging 9,744 suspicious URLs
%\footnote
%{These are reports that
(reports which contain URLs without the associated SMS text message or any relevant metadata).
%As these represent a tiny fraction of our dataset and do not report %sender IDs or text messages,
%additional useful metadata, we filter them out.
As a result, we
%we retain and further study
investigate 1,202,859 reports flagging 530,555 unique text messages. These reports originate from users located in the United Kingdom. 
The remainder of this section describes the methods we use to examine these messages. 

\subsection{Data Enrichment}
\label{subsec:data enrichment}

The 7726 user reports contain a text message, sender ID, timestamp, and URL (where available). 
We consider text separately in \S\ref{subsec:textclass} and describe next the steps we take to obtain contextual information about the remaining elements of user reports.
%phone numbers used as sender IDs, and URLs in text messages are crucial data points that we enrich to obtain additional information.

\vspace{.1cm}\noindent\textbf{Sender IDs.}
SMS texts are generally sent using a mobile number, short codes via bulk messaging services, or email addresses (incl. Apple's iMessage). Unlike email, the SMS protocol %does not include very much metadata except a
only includes sender ID and a timestamp as metadata. 
%Thus, we focus on enriching data from the phone numbers that are reported. %We cannot investigate these sender IDs without access to bulk messaging services or email providers. \marie{idk what accessing bulk messaging services has to do with sender IDs.}
%Rather, we examine the mobile numbers used to send the reported text messages.
%However, we can examine the mobile numbers used to send the text messages reported by users. 

%In particular,
We investigate the mobile numbers users report via Home Location Register (HLR) lookup~\cite{hlrlookup}. 
HLR lookup validates the queried phone number and provides the country to which the phone number belongs.
It further provides details about a mobile number -- current \mnonospace, original \mnonospace, and its current status (dead or alive). %~\cite{hlrlookup_working}. 
Current \mno is the name of a network where the phone number is currently assigned, and the original \mno is the network assigned to this telephone number range. 
These can change over time due to company name changes, mergers, acquisitions, and %a country's numbering regulator's
re-allocation of a number range~\cite{hlrlookup_api}. 

To identify the country, the current and % \mnonospace, 
original \mnonospace, and its current status, we partner with Stour Marine, who offered HLR lookups~\cite{stourmarine}.
We make these lookups against all newly used phone numbers as soon as we receive the user reports.
%\footnote{Note that we see 438,960 text messages without a sender ID. Thus, we do not have HRL lookups for these numbers.} \guillermo{This note (re: 438,960) should be moved to Section 4, somewhere. You can leave a generic note here without reporting the actual figure.}
We additionally query all phone numbers that were live the previous week. We combine these results to calculate each mobile number's lifetime and find the various \mnos scammers abuse. 

\vspace{.1cm}\noindent\textit{Survival Analysis.}
%Mobile numbers play a vital role in broadcasting scam messages. To understand the performance of the various identified scams, we study the lifetime of the mobile numbers scammers abuse to broadcast the messages that users report. 
We calculate the number of weeks 78,906 unique mobile numbers were active using the availability status we retrieve weekly from the HLR lookups. %we performed between March 14 and July 18, 2024. 

%As of our data collection last week, we see 52,477 active mobile numbers. 
We conduct survival analysis, which helps reveal patterns of activeness of mobile numbers scammer abuse. 
In particular, this technique considers intermittently unavailable data points to be ``right-censored''; we only monitor these mobile numbers until July 18, 2024, and censor those still active on that date. 
We use a Kaplan-Meier estimator~\cite{kaplan1958nonparametric} to estimate the survival function $S(t)$ from the lifetime we observe in the data. 
Intuitively, this measure illustrates the fraction of mobile numbers that become inactive after a given date.
This helps us estimate the lifetime of these numbers by using the probability of a mobile number abused by a scammer being active after $x$ weeks.

%Similar to previous research~\cite{costin_phonenumber, agarwal_usenix25}, we conduct HLR lookup on identified mobile phone numbers to identify their current and original \mno and current status. Section~\ref{subsec:senderid} provides an overview of the current and original \mnos scammers abuse to broadcast scam messages. 

\vspace{.1cm}\noindent\textbf{URLs.}
The majority of user-reported SMS texts contain URLs.
%These help identify scam text messages.
%We leverage these to identify scams by
We query all collected URLs on VirusTotal~\cite{virustotal}, an antivirus aggregator service with over 70 antivirus scanners. %, via their public API~\cite{vt_public}. 
We extract threat intelligence from VirusTotal to determine if URLs are malicious. 
We consider a URL malicious if it has been flagged as such by at least one scanner. 
%VirusTotal's public API~\cite{vt_public} to
%We collect information on our data feed, and consider that a URL is malicious if it has been flagged by at least one scanner. 
%We augment this with Google Safe Browsing's API~\cite{gsb_api} as it is possible that Google Safe Browsing may not update its results on VirusTotal.  
%
%URLs flagged by VirusTotal require a more profound understanding that helps to understand the underlying infrastructure and services abused by scammers to run SMS scams.
%We add additional metadata on each collected URL.
%We collaborate with Spamhaus~\cite{spamhaus}, WhoisXMLAPI~\cite{whoisxml}, and IPinfo~\cite{ipinfo} to query the identified domains against their APIs.

%If this is more than a year ago, we use the first date when a user reported the text message that contains the domain. 

We next describe how we further enrich our dataset of URLs with DNS telemetry, dynamic URL redirections, domain registrars, and Autonomous Systems. 

{\em Passive DNS:} Spamhaus~\cite{spamhaus} provides us with passive DNS (pDNS) API that returns the first time Spamhaus saw a domain and the IP addresses it resolved to in the last year~\cite{spamhaus_pdns}.

%We perform \texttt{WHOIS} lookup using WhoisXML API~\cite{whoisxml} to collect the domain name registrar that provided the domain.

{\em URL Shorteners:} Scammers use URL shorteners to hide the redirected malicious URL and send that in the message's text. To this end, we create a list of 27 commonly used URL shorteners and query them against the collected URLs from the text message reports.
Some of these are bespoke for a single service, like \url{wa.me} URLs that redirect to WhatsApp, while others are more generic, like \url{bit.ly}.
%We found 467 `wa.me' URLs that redirect to WhatsApp and 2094 and 689 URLs with `.sbs' and `.me' TLDs, respectively. We notice 1963 URLs with five random characters as second-level domains with `.sbs' TLD. For `.me,' we see 490 second-level domains with four random characters. As these are not URL-shortening services like Bitly, we mention them separately.

{\em Lifetime Analysis:}
The lifetime of domains helps us understand the impact of scams, as longer-running websites defraud more victims. We consider the beginning of the website to be when we first see it and the last date as per the passive DNS results.
Note that we exclude shortened URLs and other third-party domains. 

{\em Domain Registrars:} To investigate the domains scammers abuse, we first extract {\em Top-level Domains (TLDs)}.
%\footnote{we use a Python package -- {\em tld} that extracts the top-level domain (TLD) from any given URL~\cite{pypi_tld}}
We remove shortened URLs and other third-party domains (such as those using the `.me' and `.sbs' TLDs or popular domains listed among Alexa's top 10$k$ sites,
%\footnote{\href{http://s3.amazonaws.com/alexa-static/top-1m.csv.zip}{Alexa Top 1$m$ websites}}
since these domains being abused is not reflective of the registrar. %, as these are third-party services that scammers abuse indirectly. 
%We also identify 
%that are simultaneously associated with URLs abused by scammers to exploit legitimate third-party services.

We identify registration data for each domain using a WHOIS API service.
We query %{\em WHOIS} API lookups from
WhoisXMLAPI~\cite{whoisxml} at the time users report SMSs, which allows us to obtain current registration data for each domain name.
%performing Domain Registrars lookups. 
%Since we perform the lookups at the time users report SMSs, we obtain current registration data for domain names. 
%This allow us to identify threat actors to host phishing websites. %\guillermo{It is unclear how you do this and where in the results we use registration data to identify threat actors. Perhaps we are missing an experiment?}
%Domain registrars cannot act against the scammers in these cases as their customers are URL shortening services and not scammers. \guillermo{This last sentence comes a bit out of the blue... It does not relate to methodology, does it? Consider if it fits better as part of a wider discussion (e.g., Discussion section).}

{\em Autonomous Systems (ASes):} We query the IP addresses returned by Spamhaus against IPinfo's database~\cite{ipinfo} to identify the corresponding Autonomous System (AS) to the IP and the IP's geographical location~\cite{ipinfo_db}. We note that IPinfo is unable to identify the AS for 715 IP addresses.

%We use IPinfo to access their IP to AS and IP geographical location lookups to understand which AS  scammers abuse to host domains used in SMS scams and the countries where the IP addresses are based. 
%Additionally, this helps indicate possible rogue ASes that collude and provide services to scammers, also known as bulletproof hosting 

%\subsection{Analysis Methods}
%\label{subsec:analysis methods}

%We analyze user reports to identify spam and scam text messages and further investigate them to answer our research questions (\S\ref{sec:intro}).
%In this section, we explain the methods that underpin our approach. 

%\paragraph{User Report Distribution.}
\vspace{.1cm}\noindent\textbf{Timestamps.} 
%The reports we receive from our partner \mno include the timestamps when the user submits the report. As this is in
Our collected timestamps are in UTC-8. We convert the timestamp into British Summer Time (BST) which reflects local time during our collection period.%, as the data was collected between March and July. 

\subsection{Text Classification}
\label{subsec:textclass}

We perform text analysis to characterize user reports, classify scam messages, and identify the lures that scammers use.  

\vspace{.1cm}\noindent\textbf{User Report Characterization.}
We categorize the different types of user reports to whitelist messages that are not scams. 
%Towards this end, we identify URLs, suspicious calls, and text messages reported by users. 
%
%For suspicious calls, we check if the report only has the words `call' or `voice' or reports a number without any text. Additionally, if the report starts with `missed call:' or `voicemail' or has either of these two keywords without a URL in the report, we categorize these user reports as suspicious calls. Next, 
In particular, 
we look into the text messages and search for specific keywords in the text as described in Table~\ref{tab:keywords} to identify straightforward cases: (1) text messages containing one-time passwords (OTPs) and (2) spam text messages. 
OTP messages are occasionally reported to 7726 when they are unintentionally delivered to the wrong recipient,\footnote{For example, a sender may mistype the intended recipient's phone number, resulting in the OTP being sent to an unrelated individual. This can lead the recipient to mistakenly perceive the message as spam or a phishing attempt, prompting them to report it to 7726.} 
but they are easy to flag. 
Likewise, spam messages are often reported to 7726. 
These messages are easier to characterize than scam messages because they follow more predictable patterns and adhere to explicit opt-out mechanisms. 
After filtering out the more straightforward cases, we next outline the mechanism employed to further identify and analyze scam messages.
%Fig.~\ref{fig:user_report_distribution} illustrates our approach to characterize all user reports into various categories. 

\begin{table*}[!ht]
    \centering
    \scalebox{0.9}{
    \resizebox{\linewidth}{!}{%
\begin{tabular}{lr}
    \hline
    \multicolumn{1}{c}{\bf Classification} & \multicolumn{1}{c}{\bf Keywords Used}\\
    \hline
    \hline
    Spam Text Messages & STOP, optout, opt out, opt-out, won the draw, claim your prize and offer ends.\\
    Text Messages with One Time Password (OTP) & otp and verification code \\
    \hline
\end{tabular}
}
    }
    \caption{Keywords used to differentiate suspicious text messages from spam and messages with OTPs.}
    \label{tab:keywords}
\end{table*}

\begin{comment}
\begin{figure}[!ht]
\centering
\resizebox{\linewidth}{!}{%
\begin{circuitikz}
\tikzstyle{every node}=[font=\LARGE]
\draw [ line width=0.7pt ] (5.25,14.25) rectangle  node {\LARGE 7726 User Reports} (11.25,12.75);
\draw [ line width=0.7pt ] (2,10.75) rectangle  node {\LARGE URLs Only} (5.5,9.5);
\draw [ line width=0.7pt ] (6,10.75) rectangle  node {\LARGE Suspicious Calls} (10.5,9.5);
\draw [ line width=0.7pt ] (11,10.75) rectangle  node {\LARGE Text Messages} (15.25,9.5);
\draw [line width=0.7pt, ->, >=Stealth] (8.25,12.75) -- (4,11);
\draw [line width=0.7pt, ->, >=Stealth] (8.25,12.75) -- (8.25,11);
\draw [line width=0.7pt, ->, >=Stealth] (8.25,12.75) -- (12.75,11);
\draw [ line width=0.7pt ] (3,6.75) rectangle  node {\LARGE Spam Text Messages} (9,5.5);
\draw [ line width=0.7pt ] (9.75,6.75) rectangle  node {\LARGE Suspicious Text Messages} (16.75,5.5);
\draw [ line width=0.7pt ] (18.5,6.75) rectangle  node {\LARGE Text Messages with OTPs} (25.75,5.5);
\draw [line width=0.7pt, ->, >=Stealth] (12.75,9.5) -- (6.25,7);
\draw [line width=0.7pt, ->, >=Stealth] (12.75,9.5) -- (12.75,7);
\draw [line width=0.7pt, ->, >=Stealth] (12.75,9.5) -- (22.25,7);
\end{circuitikz}
}%
\caption{Breakdown of user reports into different identified categories.}
\label{fig:user_report_distribution}
\end{figure}
\end{comment}

\vspace{.1cm}\noindent\textbf{Scam Text Message Identification.}
We investigate the collected suspicious text messages to identify spam and scam text messages. To this end, we create our prompt (cf. Appendix~\ref{appendix:openai_prompt}) and query the unique suspicious text messages using OpenAI's API GPT-4o model. We select GPT-4o as this was the best available model during our research period.  %Fig.~\ref{fig:sus_text_message_distribution} visualizes our steps to identify scam text messages and classify them into different scam types using OpenAI.

First, we run all unique suspicious text messages with our prompt via OpenAI's model to categorize the text messages into spam. % and classify the other suspicious text messages without URLs into various scam types. 
Next, we query the URLs from the suspicious text messages for antivirus detection on VirusTotal and query all the identified malicious domains against the domains in the text of all unique suspicious text messages with URLs. 
Lastly, we classify suspicious text messages with and without URLs using OpenAI's GPT-4o model into various scam types. 
In particular, we identify the six known SMS scam types in the UK introduced in Section~\ref{sec:background}, i.e.: (1) Wrong Number, (2) Hi Mum/Dad, (3) Delivery, (4) Banking, (5) Telecom, and (6) Government~\cite{agarwal_imc24}.  
The remaining suspicious text messages with URLs where VirusTotal does not flag the URL are  `Unknown.' 

To our surprise, the initial results for scam type classification had almost 50\% of the scam text messages marked as `Others'. To this end, we update our prompt (cf. Appendix~\ref{appendix:openai_prompt}) and ask OpenAI to suggest a category if the initial classification is `Others.'

\vspace{.1cm}\noindent\textbf{Complexity Analysis.}
Gunning Fog Index helps explain the educational level required to understand the scam text message the Gunning fog index~\cite{gunning1969fog}.
%To this end, we perform tokenization using Python's NLTK library\footnote{\href{https://www.nltk.org/}{Python NLTK library}} to perform natural language processing on the SMS text.
The scam text messages collected in our dataset are in English, as they target users in the UK. We tokenize the text of the SMS messages reported by users and remove the stop words. Stop words are common words that do not provide meaningful information about the topic. 
%For example, `the,' `and,' and `I.' We use NLTK's stopword list for this purpose.
To calculate the complex words, we use the conventional criteria of the Gunning Fog Index and filter out the words with less than three syllables. We additionally remove proper nouns, such as brand names being impersonated. %, we also remove them before calculating the complex words.
The formula to calculate the Gunning Fog Index is as follows: 

{
\small
\[
\text{GFI} = 0. 4^{\ast}\left(\left({words\over sentence}\right)+100\left({complex words\over words}\right)\right)
\]
}

%where Words - Total number of words in the scam text message. Sentences - Total number of sentences in the scam text message. 
%Complex Words - Words with three or more syllables, excluding proper nouns, familiar jargon, or compound words. 
where words and sentences are counts as normally defined and complex words are words with three or more syllables, excluding proper nouns, familiar jargon, or compound words. 
%\S\ref{subsec:text_analysis} presents the results of the Gunning Fog Index for identified scam texts. 

\vspace{.1cm}\noindent\textbf{Scam Lure Analysis.}
We want to understand the various lures that scammers use to deceive victims into taking actions mentioned in a scam message. Towards this end, we adopt the ontology of lures from Stajano and Wilson~\cite{StajanoFrank2011_principles}. 
The n-gram analysis for multiple scam types with thousands of messages is not feasible to do manually. Thus, we use OpenAI's GPT-4o to categorize our messages into the different lures using the prompt listed in Appendix~\ref{appendix:openai_prompt_lures}.
%is not feasible due to a manual approach. Therefore, we utilize large language models (LLMs). Using OpenAI's GPT-4o model, we provide a prompt (cf. Appendix~\ref{appendix:openai_prompt_lures}) that returns the various lures scammers use to deceive victims into identified scams in \S\ref{subsec:text_analysis}. 

\subsection{Evaluation}

%\paragraph{Evaluation of OpenAI Classification.}

%We utilize OpenAI's GPT-4o large-language model (LLM) to identify and classify scam text. 
%\guillermo{I've removed a sentence here (see commented it out) because you finish the previous section saying the same thing, and you have repeatedly mentioned that you use GPT4 throughout it. 
%\guillermo{Also, review below how you refer to OpenAI (e.g., "The ... indicates that OpenAI performs very well"). I would instead talk about your methodology. Your wording seems to downplay the complexity behind your method. It gives the impression that the study can be easily done using an OpenAI prompt, and in reality there is way more to it.}
To evaluate the performance of our methodology, % and the OpenAI model, 
we extract 384 random texts from our dataset, ensuring (95\% confidence) that the random sample is representative of the complete dataset.
We manually label them as an OTP, spam, or scam. Next, we extract another random sample of 384 scam texts and manually classify them into seven scam types (including other) along with the scammer lures used in each scam text.
We use human-labeled texts as ground truth and calculate the inter-rater reliability (IRR) between ground truth and OpenAI's annotation. 
We use Cohen's $\kappa$~\cite{cohen1960coefficient}, a standard metric for IRR.
There is near-perfect agreement for OTP and scam types and substantial agreement for spam and lures (see Table~\ref{tab:irr_eval} for Cohen's $\kappa$ coefficient and F1-score). 

\begin{table}[!ht]
    \centering
    \scalebox{0.9}{
    \begin{tabular}{lrlr}
    \hline
    \multicolumn{1}{c}{\bf Category} & \multicolumn{1}{c}{\bf $\kappa$ coeff.} &\multicolumn{1}{c}{\bf Agreement level} &\multicolumn{1}{c}{\bf F1-score}\\
    \hline
    \hline
    One Time Password (OTP) & 0.81 & Near-perfect/ Strong & 91\%\\
    Spam & 0.76 & Substantial/Moderate & 88\%\\
    Known Scam Types & 0.84 & Near-perfect/ Strong & 89\%\\
    New Scam Types & 0.71 & Substantial/Moderate & 80\%\\
    Lures & 0.73 & Substantial/Moderate & 78\%\\
\hline
\end{tabular}
    }
    \caption{Cohen's $\kappa$ Coefficient for evaluating the inter-rater reliability between ground truth and our methodological framework using OpenAI ($n=384$).}
    \label{tab:irr_eval}
\end{table}

%\guillermo{I would refrain from qualifying if the performance is ``very well'' or instead if it is ``only'' this and that. Be objective -- mention the exact value and use the wording you have for the agreement levels (e.g., "substantial" agreement level). That terminology has been established and by using it, you avoid giving the impression that you are giving your opinion.}
%We find that OpenAI is able to identify our known scam types 88\% of the time and spam 93\%. 
%The $\kappa$ in Table~\ref{tab:irr_eval} indicates that OpenAI performs very well in identifying known scam types.

%While OpenAI has a near-perfect agreement for known scam types,  
In addition to prompting AI to label scams based on existing scam types, we also asked AI to provide a new scam category rather than ``other." However, this new scam category suggestion has only a moderate agreement level (see Table~\ref{tab:irr_eval}). We also find that its category recommendations are inconsistent for the same scam type. 
%when we ask it to provide a new scam category if the text message does not belong to one of the six known categories, 
%This new scam category suggestion is only 72.4\% correct.
For example, in the case of `job scam' texts, OpenAI provides a variety of new scam categories --- `job/recruitment', `employment', `joboffer', `jobopportunity', and in 12.3\% cases, it even returns no new category or `Unclear.' 
The inconsistency here justifies our two-stage approach -- 
%This indicates that OpenAI is inconsistent in naming new scam types and cannot always provide a new category, which justifies our two-stage approach --- 
one with a fine-grained classification under a closed-world of scam types and another with an open-world assumption refining those in the ``others'' category.

%OpenAI correctly identifies the lures that scammers use to deceive victims for 82\% of the sample scam texts~\cite{StajanoFrank2011_principles}. 
%While it misclassifies only 5\% of the scam texts, it misses one lure principle in 12\% of cases.

\section{Results}
\label{sec:analysis}

We observe 1,202,859 SMS reports over four months. 
This section presents the distribution of these user reports over time and the scam text messages we identify. 
We study existing scam types and the infrastructure scammers abuse to conduct them, including the \mnos that they abuse. 
Lastly, we highlight the lures scammers use and examine the readability of the scam text messages.
% and the broad types of reports that we identified.
%\guillermo{Give a concise overview of the other subsections}

\subsection{User Reports Over Time}
\label{subsubsec:reports_over_time}

\begin{figure}[!ht]
    \centering
    \includegraphics[width=.8\linewidth]{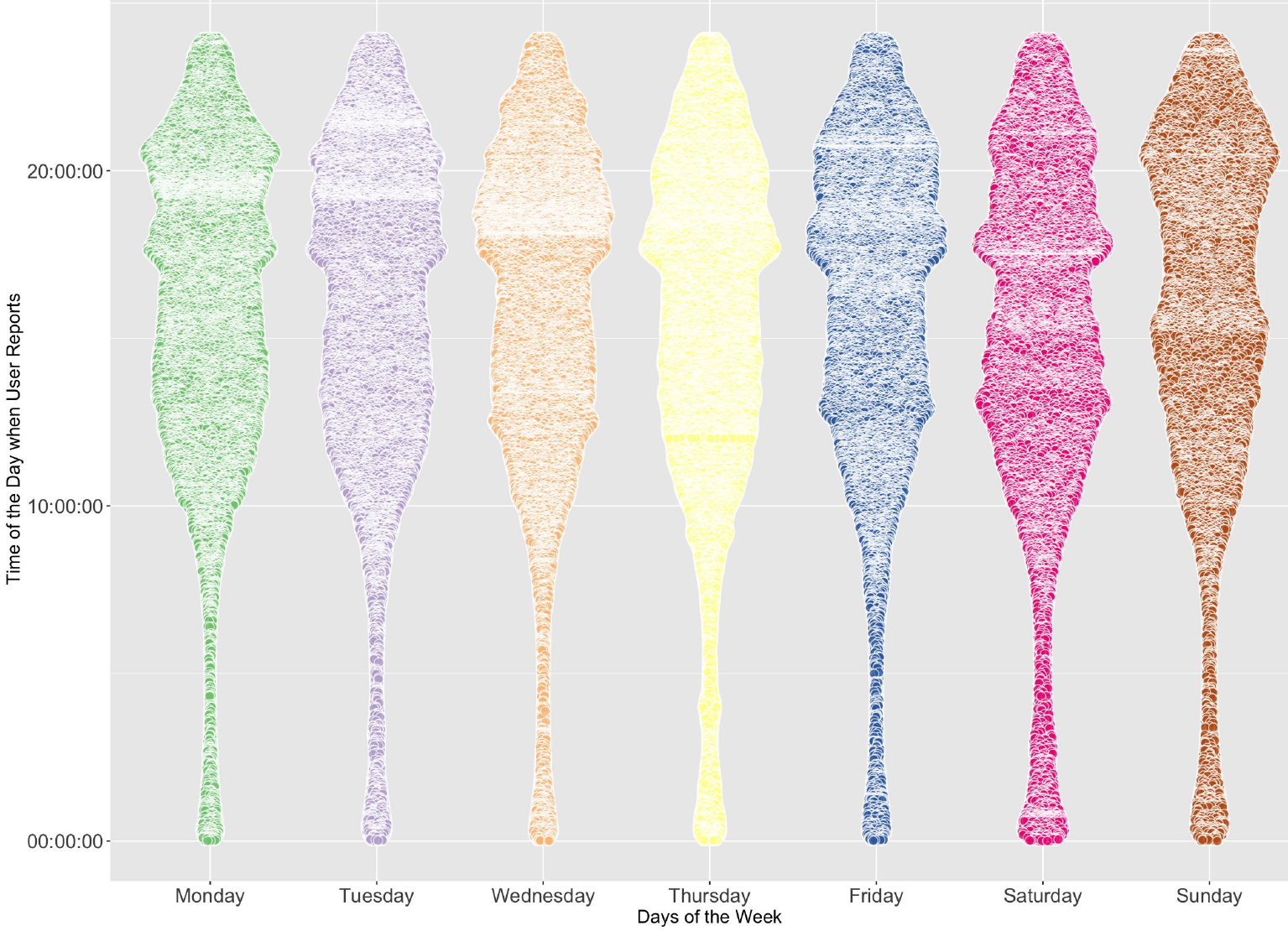}
    \caption{Time of the day per week when users report suspicious text messages ($n=1,202,859$). The pair-wise two-sample KS test is significant with $p<0.05$.}
    \label{fig:time_per_day_report}
\end{figure}

Table~\ref{tab:reports_weekly} presents the weekly distribution of user reports along with associated sender IDs.\footnote{The identity of the user submitting a report remains anonymous and should not be confused with sender IDs.} 
The average number of reports submitted by users weekly is 66.8$k$, with a median of 64.9$k$. 
From these reports, we see a weekly average and median of 11.8$k$ unique sender IDs. 
%Ofcom, the UK telecom regulator, found that only 15\% of surveyed people were aware of the 7726 reporting number~\cite{ofcom_userexperience}. \guillermo{So what re: Ofcom? I understand you want to put things in context to suggest that in the wild there may be more, but the discussion is incomplete. TBH, I would kill this sentence and stick to the next section you also comment statistics from ofcom.}
Fig.~\ref{fig:time_per_day_report} shows the time of the day per week when users report suspicious messages in BST. Most users report between 10:00 - 21:00 BST (with medians: Mon - 16:10:28, Tues - 16:15:00, Wed - 16:22:25, Thurs - 16:05:34, Fri - 16:26:41, Sat - 15:16:26, Sun - 16:11:25).
%We perform a Kolmogorov-Smirnov (KS) test to examine the daily distribution of user reports and 
%find that the daily distribution of users' reports differs ($p<0.05$).
%The p-value for all combinations is significant, $p<0.05$, indicating that the 
%\guillermo{What is the implication of having different daily distribution of user reports? This sub-section is missing the last mile, but perhaps you can argue something like the following (please, update based on your expertise working with StopScamsUK or kill if you see it does not fly):}
This pattern aligns with the opportunistic nature of scam campaigns, which are often driven by specific events or timing rather than a steady stream of messages. In contrast, spam campaigns may follow consistent schedules or promotional cycles.
%The daily distribution of users' reports is different. The p-value for the two-sample Kolmogorov-Smirnov (KS) test for all combinations is significant ($p<0.05$). \marie{consider rewording this \url{https://psych.uw.edu/storage/writing_center/stats.pdf}}

\begin{table}[!ht]
    \centering
    \scalebox{0.9}{
    %\resizebox{\columnwidth}{!}{%
\begin{tabular}{lrrrr}
    \hline
    \multicolumn{1}{c}{\bf Dates} & \multicolumn{2}{c}{\bf Reports ($k$)} & \multicolumn{2}{c}{\bf Sender IDs ($k$)}\\
    \multicolumn{1}{c}{\bf (2024)} & \multicolumn{1}{c}{\bf Total} & \multicolumn{1}{c}{\bf Unique} & \multicolumn{1}{c}{\bf Total} & \multicolumn{1}{c}{\bf Unique}\\
    \hline
    \hline
    Mar 8 - 14 & 73.3 & 36.2 & 45.9 & 13.2\\
    Mar 14 - 21 & 71.9 & 37.7 & 43.9 & 11.9\\
    Mar 21 - 28 & 70.8 & 39.2 & 43.1 & 12.9\\
    Mar 28 - Apr 4 & 59 & 33.8 & 36.5 & 11.7\\ 
    Apr 4 - 11 & 62.1 & 34.8 & 38 & 11.9\\
    Apr 11 - 18 & 61.4 & 34.3 & 37.9 & 12\\
    Apr 18 - 25 & 62.8 & 34.5 & 39.1 & 12.9\\
    Apr 25 - May 2 & 72.9 & 34.1 & 47 & 11.9\\
    May 2 - 9 & 66.9 & 32.5 & 43.9 & 11\\
    May 9 - 16 & 57.8 & 33.1 & 36.1 & 11.6\\
    May 16 - 23 & 63.6 & 34.9 & 40.9 & 11.6\\
    May 23 - 30 & 69.9 & 41.4 & 45.8 & 12\\
    May 30 - Jun 6 & 62.9 & 34 & 40.3 & 11.4\\
    Jun 6 - 13 & 65.3 & 33.6 & 42.6 & 11.5\\
    Jun 13 - 20 & 63 & 28.9 & 40.8 & 10.6\\
    Jun 20 - 27 & 64.5 & 28.6 & 41.6 & 10.3\\ 
    Jun 27 - Jul 4 & 74.3 & 31.7 & 48.9 & 11.5\\
    Jul 4 - 11 & 80 & 33.7 & 51.4 & 12.4\\
    \hline
\end{tabular}
%}
    }
    \caption{Distribution of all text message reports ($n=1,202,859$) received weekly.}
    \label{tab:reports_weekly}
\end{table}

\vspace{.1cm}\noindent\textbf{Takeaway.}
We see that users mostly submit reports between 10:00 and 21:00 BST daily. 
Prior work found that scammers interact with Hi mum/dad scam victims during 10:00-15:00 UK time~\cite{agarwal_usenix25}, indicating that users report SMS messages or calls without significant delays. 

\begin{comment}
Users report suspicious phone calls and SMS messages to 7726. We cannot investigate suspicious calls as the reports only contain a mobile phone number; we primarily focus on the SMS text messages reported by users. Fig~\ref{fig:user_report_distribution} shows the overview of identified categories of reports users submit. 

We categorize these reports into three broad categories: (1) text messages, (2) suspicious calls, and (3) URLs only. Table~\ref{tab:category_report_distribution} shows each category's total and unique reports. Ofcom reports that only 6\% users in the UK have used 7726 to report a suspicious text, and 4\% to report a suspicious call~\cite{ofcom_userexperience}. Users report suspicious calls by directly reporting the phone numbers or having keywords mentioned in \S\ref{subsec:analysis methods}. These reports do not include any URLs or an additional phone number. For example, 

\begin{displayquote}
Missed call: this person/number called at 09:23 on 14th Mar but left no message.

Voicemail message received 13:49 on 10th Mar. Duration 18 sec. Click or call <shortcode> to hear this message at your normal call rate.
\end{displayquote}

URLs only refer to reports in which users report only a URL without any text. We categorize all remaining reports as text messages. 

\begin{table}[!ht]
    \centering
    \input{tables/category_report_distribution}
    \caption{Distribution of all user reports ($n=1,349,039$) into text messages, suspicious calls, and URLs only.}
    \label{tab:category_report_distribution}
\end{table}
\end{comment}

\subsection{Characterization of User Reports}
\label{subsec:characterization_user_reports}

To answer \ref{rq1}, we study the type of suspicious text messages in \S\ref{subsec:Suspicious Text Message Classification}. We characterize the type of scams in \S\ref{subsubsec:openai_class} to answer \ref{rq2}.
%\guillermo{I feel RQs are a descontextualized and it may be a good idea using them to build the scaffolding of Section 4. Please, review above and replicate for other RQs. This exercise may let you assess if the structure of this section is okay and/or if RQs make sense.}

\subsubsection{Suspicious Text Message Classification}
\label{subsec:Suspicious Text Message Classification}

The first step of our text classification method focuses on identifying and whitelisting clear cases of fraudulent and non-fraudulent messages. 
Table~\ref{tab:suspicious_sms_distribution} presents the results of our initial report characterization, where we find 7,143 (1.35\%) text messages delivering OTPs, 186,325 (35.12\%) spam messages, and 213,659 (40.27\%) scam messages. 
%First, we identify 7,143 text messages delivering OTPs. 
This result indicates that 7726 user-reported messages contain a significant amount of non-fraudulent messages, prompting subsequent steps in our methodology to filter out irrelevant content and further classify types of fraud.
The results of the subsequent steps in our methodology are presented in \S\ref{subsubsec:openai_class}.
We now delve deeper into these results, providing detailed discussions and examples of the different types of messages observed and reasons that may drive users to report non-fraudulent messages.

\begin{table}[!ht]
    \centering
    \scalebox{0.9}{
    %\resizebox{\columnwidth}{!}{%
\begin{tabular}{l|rr|rr}
    \hline
    \multicolumn{1}{c|}{\bf Type} & \multicolumn{2}{c}{\bf Unique Text Messages} & \multicolumn{2}{|c}{\bf Total}\\
    & \multicolumn{1}{c}{\bf w/ URLs} & \multicolumn{1}{c}{\bf w/o URLs} & \multicolumn{1}{|c}{\bf (\#)} & \multicolumn{1}{c}{\bf (\%)}\\
    \hline
    \hline
    OTPs & - & 7,143 & 7,143 & 1.35\\
    Spam & 164,826 & 21,499 & 186,325 & 35.12\\
    Scam & 46,635 & 167,024 & 213,659 & 40.27\\    
    \hline
    Unknown & 123,428 & - & 123,428 & 23.26\\
    \hline
    \hline
    Total & 334,889 & 195,666 & 530,555 & 100\\
    \hline
    %Spam (from Table~\ref{tab:sms_distribution}) & 81,878 & 12,252\\
    %\hline
    % \hline
    % Spam & 92,195\\
    % \hline
    % Unknown & 123,428\\
    % \hline
\end{tabular}
%}
    }
    \caption{Categorization of all unique text messages 
    ($n=530,555$) into spam and scam, both with and without URLs.}
    \label{tab:suspicious_sms_distribution}
\end{table}

\vspace{.1cm}
\noindent{\em Non-fraudulent messages.} 
We identify 36.47\% of the messages as non-fraudulent (7,143 OTPs, 186,325 Spam). 
While most of these are spam (35.12\%), we see a small fraction of unique reports containing OTPs (1.35\%). 
For example,

\begin{displayquote}
\small
$<$brand name$>$: 1041 is your verification code. It expires in 15 minutes. Don't share this with anyone.
\end{displayquote}

\noindent Examples of spam messages include:

\begin{displayquote}
\small
Join now to receive a $<$brand$>$ 100 FS plus up to 2000 GBP and enjoy weekly bonuses [URL]  OptOut: [URL].
\end{displayquote}

These messages are generally unsolicited marketing messages sent by companies or unknown senders. 
Spam is often annoying and reduces the availability of messages on a user's mobile phone, which can prompt users to report them to 7726.
%We also attribute the reporting of non-fraudulent messages to errors. 
Some of these reports might be erroneous.
As reported by Ofcom, %the UK regulator,
the built-in reporting function on users' mobile phones was the most used channel to report suspicious messages~\cite{ofcom_userexperience}.
The user interface has `delete' and `delete and report' buttons next to each other, which could confuse users and send their reports to 7726 instead of simply deleting the text message.

\vspace{.1cm}
\noindent{\em Fraudulent messages.} 
We identify 40.27\% (213,659) of the unique messages as fraudulent. 
Out of these, we see 46,635 with URLs and 167,024 without URLs. 
Overall, users report 334,889 messages with URLs, which means that most of the messages reported with URLs are either spam or the URL is not flagged by an AV vendor on VirusTotal. 
VirusTotal only flags 24,546 (44.77\%) of our unique URLs as suspicious (9,285) or malicious (20,482). 
This indicates that \mnos are more effective in blocking scam texts with URLs. 
Alternatively, it could also mean that the scammers are shifting towards scams without initial URLs. 

While URLs are more common in spam than in scams, they can play a crucial role in identifying fraudulent messages. 
The following illustrates a smishing attack, featuring a seemingly benign SMS, where the only distinguishing factor between a legitimate and malicious message lies in the link's behavior: 

\begin{displayquote}
\small
$<$Brand name$>$: Hi, unfortunately you have missed your delivery. Please visit [URL] to schedule a redelivery.  
\end{displayquote}

\vspace{.1cm}
\noindent{\em Unknowns.} 
We label the remaining 123,428 (23.3\%) unique suspicious text messages with URLs as `Unknown'. 
Considering that these messages have been reported by users and flagged as suspicious, we see them as potentially malicious rather than benign due to the various evasion techniques deployed by phishing websites~\cite{doupe_usenix20} and the low recall of AV vendors in detecting phishing websites~\cite{peng_imc19,srinivasan2016understanding}.

\subsubsection{Scam Type Classification}
\label{subsubsec:openai_class}

\begin{figure*}[!ht]
    \centering
    \begin{subfigure}{0.485\linewidth}
    \centering
    \includegraphics[width=\linewidth]{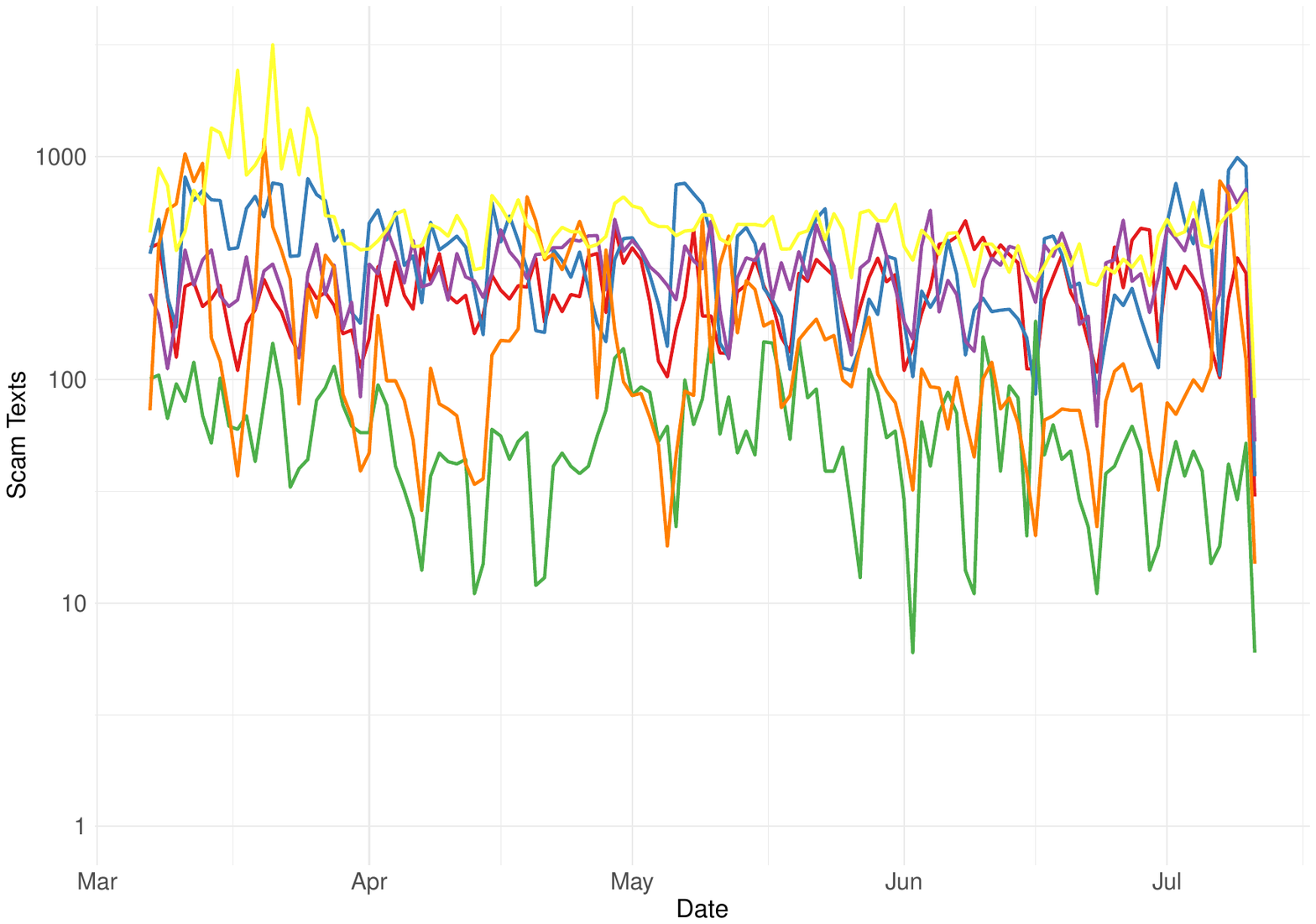}
    \caption{Distribution of known scams' texts over time ($n=222,143$). Yellow: Wrong number, Red: Banking, Blue: Delivery/Parcel, Green: Government, Purple: Hey Mum/Dad, and Orange: Telecom scams.}
    \label{fig:temporal_distribution_known_scams}
    \end{subfigure}
    \begin{subfigure}{0.01\linewidth}
    \caption*{}
    \end{subfigure}
    \begin{subfigure}{0.485\linewidth}
    \centering
    \includegraphics[width=\linewidth]{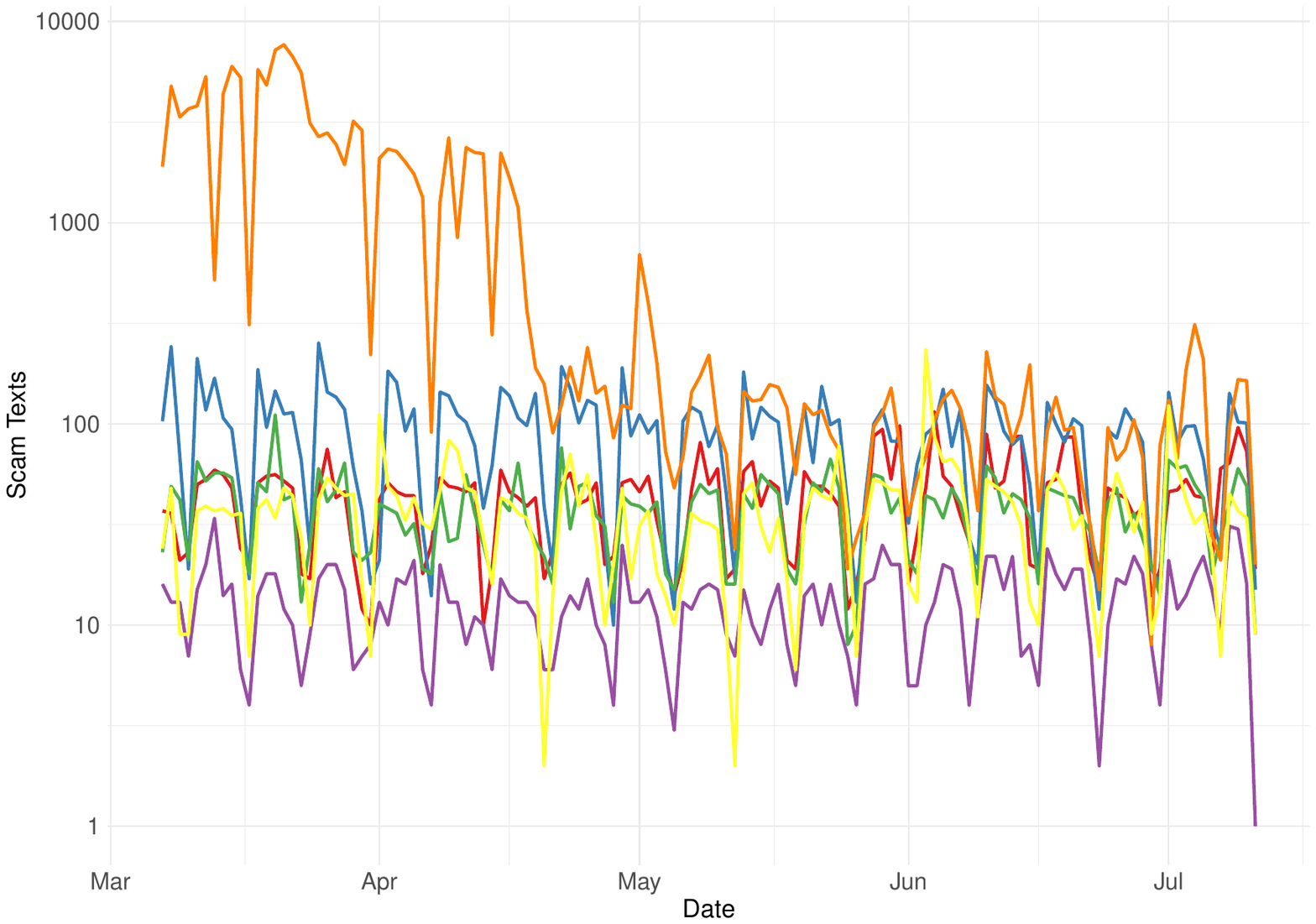}
    \caption{Distribution of new scams' texts over time ($n=163,808$). Orange: Job, Blue: Debt, Red: Appointment, Green: Finance, Yellow:  Utility, and Purple: Insurance scams.}
    \label{fig:temporal_distribution_new_scams}
    \end{subfigure}
    \caption{Distribution of known and new scam-type messages over time. Y-axis is on log scale. Colors represent scam types.}
    \label{fig:amount distribution}
\end{figure*}

We classify scam text messages into six known SMS scam types~\cite{agarwal_imc24} and identify six new scam types using our methodological process described in \S\ref{subsec:textclass}. 
Table~\ref{tab:scam_types} outlines messages by category.
Out of the 119,398 scam texts overall, we discover that the most popular scam type is the Wrong Number scam (16.36\%), followed by Banking (9.14\%) and Delivery/Parcel (6.81\%) scams. %We also notice that the following two scams by volume are Job and Debt-related scams, which are new scam categories. 
We plot the distribution of the these scams over time in Fig.~\ref{fig:temporal_distribution_known_scams}.

\begin{table}[!ht]
    \centering
    \resizebox{\columnwidth}{!}{%
\begin{tabular}{l|l|rr|rr}
    \hline
     & \multicolumn{1}{c}{\bf Category} & \multicolumn{2}{| c}{\bf Unique Scam Texts} & \multicolumn{2}{|c}{\bf Total}\\
     & & \multicolumn{1}{c}{\bf w/o URL} & \multicolumn{1}{c}{\bf w/ URL} & \multicolumn{1}{|c}{\bf (\#)} & \multicolumn{1}{c}{\bf (\%)}\\
    \hline
    \hline
    \multirow{6}*{Known \cite{agarwal_imc24}} & Wrong Number & 34,863 & 86 & 34,949 & 16.36\\
    & Banking & 14,744 & 4,784 & 19,528 & 9.14\\
    & Delivery/Parcel & 1,834 & 12,725 & 14,559 & 6.81\\
    & Hi Mum/Dad & 6,604 & 190 & 6,794 & 3.18\\
    & Telecom & 3,743 & 2,083 & 5,826 & 2.73\\
    & Government & 1,880 & 2,229 & 4,109 & 1.92\\
    \hline
    % Others & 103,356 & 24,538 & 127,894\\  
    % \hline
    \multirow{6}*{New} & Job & 10,407 & 528 & 10,935 & 5.12\\
    & Debt & 7,188 & 3,185 & 10,373 & 4.85\\
    & Appointment & 4,087 & 458 & 4,545 & 2.13\\
    & Finance & 2,800 & 1,018 & 3,818 & 1.79\\
    & Utility & 1,253 & 1,355 & 2,608 & 1.22\\
    & Insurance & 1,201 & 153 & 1,354 & 0.6\\
    \hline
    \hline
    %\multirow{1}*{Unidentified} 
    & Sub-Total & 90,604 & 28,794 & 119,398 & 55.88\\ %77,117\\
    \hline
    & Others & 76,420 & 17,841 & 94,261 & 44.12\\ %77,117\\
    \hline
    \hline
    & \multicolumn{1}{l|}{Total} &  167,024 & 46,635 & 213,659 & 100\\
    \hline
\end{tabular}
}
    \caption{Identified categories of scam text messages ($n=213,659$), both with and without URLs.}%\guillermo{Does it fit in one column, even if it means scaling to .9 or .85?}}
    \label{tab:scam_types}
\end{table}

From the new scam categories, we find that job-related scams (5.12\%) are the most reported category, followed by debt (4.85\%) and appointment-related scams (2.13\%).
%Utility scams are texts in which scammers impersonate a utility company and ask users to pay their bills or enter other details. Finance-related scams include a text asking users to verify their finance information such as:
%\begin{displayquote}
%Please ensure we have the correct student finance %information for your April payment by visiting: [URL]
%\end{displayquote}
Job-related scams lure victims by providing fake employment opportunities~\cite{which_scams}. Debt scams are text messages that %provide a URL
ask users to pay an outstanding debt balance. 
%Debt, Finance, Utility, and Insurance-related scams deceive victims by impersonating brands and asking users to pay debt/bills that are overdue by clicking on a URL, call, or reply to a text. 
Appointment scams are similar to appointment reminders, asking users to call or text back on the phone number but then luring victims into transferring funds or stealing financial or personal details~\cite{appointment_scam,fakeappointment_scam}. Example of job-related scams is (cf. Appendix~\ref{appendix:other_scam_types} for other types):

\begin{displayquote}
\small
Hi, I'm $<$First Name$>$ from $<$Fake HR$>$. I'd like to introduce you to a employment opportunity here, can I share some detail with you?
\end{displayquote}

Some scam texts contain a malicious URL and are known as smishing or SMS phishing.
These messages contain a malicious URL enticing a user to click; 
%asking the user to click on it,
this redirects to a phishing page impersonating a brand which deceives victims into providing personal or financial details. Unsurprisingly, Delivery/Parcel (12,725) and Banking (4,784) impersonation scams are the top two categories for URL-based scams (Table~\ref{tab:scam_types}). 
From the new categories, for URL-based scams, we find Debt (3,185) and  Utility (1,355) scams to be the top two. Utility scams impersonate a utility company, such as gas or water, and aim to steal users' details and banking credentials~\cite{utility_scam}.
%This contrasts with scams without URLs which feature Wrong Number scams followed by banking scams. %followed by government and telecom scams.
Here is an example Banking scam text with a URL where scammers try to lure victims into providing  credentials:

\begin{displayquote}
\small
$<$Bank Name$>$: A scheduled payment to $<$Brand Name$>$ has been made, please verify your credentials via: [URL]
\end{displayquote}

% From the new scam categories, we find that job-related scams are the most reported category, followed by debt and appointment-related scams.
% %Utility scams are texts in which scammers impersonate a utility company and ask users to pay their bills or enter other details. Finance-related scams include a text asking users to verify their finance information such as:
% %\begin{displayquote}
% %Please ensure we have the correct student finance %information for your April payment by visiting: [URL]
% %\end{displayquote}
% Job-related scams lure victims by providing fake employment opportunities~\cite{which_scams}. Debt scams are text messages that %provide a URL
% ask users to pay an outstanding debt balance. 
% %Debt, Finance, Utility, and Insurance-related scams deceive victims by impersonating brands and asking users to pay debt/bills that are overdue by clicking on a URL, call, or reply to a text. 
% Appointment scams are similar to appointment reminders, asking users to call or text back on the phone number but then luring victims into transferring funds or stealing financial or personal details~\cite{appointment_scam,fakeappointment_scam}. Examples of job and debt-related scams are (cf. Appendix~\ref{appendix:other_scam_types} for other types):

% \begin{displayquote}
% Hi, I'm Jenny Marsden from Bright HR. I'd like to introduce you to a employment opportunity here, can I share some detail with you?

% An Enforcement Agent has been scheduled to attend your property, call <name> NOW ON <phone number> to prevent this action. Quote ref <reference number>
% \end{displayquote}

Conversational scams, on the other hand, are text messages asking users to interact with the scammer directly via text. For example, in Hi Mum/Dad scams or Wrong Number scams, scammers send an initial text message and deceive the potential victim into replying/initiating the conversation on the same or a new mobile phone number. For example, 

\begin{displayquote}
\small
Now then mate you well - just a txt to say hello buddy . X Sending love - and sorry I've not been in contact x

%Just a quick heads up mum - from my tablet - to let you know about my recent contract upgrade. Text me on the new number <mobile number> \& save it cheers x
\end{displayquote}

We do not expect conversational scams to contain URLs. However, we see a handful of these (86 Wrong Number, 190 Hi Mum/Dad) with URLs,
%(86 Wrong Number + 190 Hi Mum/Dad),
overwhelmingly from scammers attempting to move the conversation to online messaging platforms, e.g., WhatsApp. 
%Our results show that the identified scams from user reports that use a URL are comparatively lower than the scam text messages without a URL.
%In conversational scams such as Wrong Number and Hi Mum/Dad scams, 
%Table~\ref{tab:scam_types} also presents the distribution of scam types for scam text messages with and without URLs. 
Contrary to URL-based scams, scams without URLs feature Wrong Number scams (34,863), followed by Banking scams. %followed by government and telecom scams.
%Unsurprisingly, Wrong Number and Hi Mum/Dad scams are in the top three scams for messages without URLs as 
These are more difficult to identify than a malicious URL in a smishing text. While Hi Mum/Dad scams are a well-known authorized push payment (APP) fraud which cause significant financial loss to victims~\cite{agarwal_usenix25}, these are actively blocked by \mnos in the UK~\cite{agarwal_imc24} and thus lower in ranking by volume (6,604). 
Investigating banking scam messages without a URL, we find that these messages mention an OTP or a transaction and ask the victim to call the provided phone number. For example, 

\begin{displayquote}
\small
$<$Bank name$>$ Bank: Transaction $<$brand name$>$ £442.26 on 19/05/2024 15:30pm .Your code is 446228. If this was NOT you call us on $<$phone number$>$
\end{displayquote}

This indicates a rise in text and call-back scams. Here, scammers lure victims through text messages by asking them to text or call the phone number provided in the SMS text. These messages often include a OTP or mention a fake transaction, instructing recipients to call or text a provided phone number if they did not request the OTP or initiate a transaction. 
%Our study does not include directly engaging with phone numbers provided, such as calling or texting;
We do not directly engage with phone numbers provided, either by texting or calling; telephony honeypots present a potential avenue for deeper investigation~\cite{miramirkhani2016_techscam} discussed in \S\ref{sec:discussion}.

%For the new scam categories for non-URL based scams, Table~\ref{tab:scam_types} 
We note new scam categories have higher volumes for non-URL-based scams, particularly Job scams (10,407) and Debt-based scams (7,188)%as the top two scams reported by users
~\cite{debt_scamwithouturl,job_scam,which_scams}. Job scams exploit users' vulnerability by contacting them about unrealistic job opportunities, offering high salaries to lure them into a scam~\cite{job_scamwhich, ftc_jobscams}. We plot the distribution of the new scam types over time in Fig.~\ref{fig:temporal_distribution_new_scams}.

%For example,
% \begin{displayquote}
% <Bank name>OTP: A new payee was added 04.06.24 <firstname lastname> if this was not done by you contact us on <phone number>    

% <Bank name> : YOUR OTP IS 434467 TO <brand name> - £754.54 at  14:04:11. CALL US  URGENTLY ON <phone number>. REF : BC924TR
% \end{displayquote}

%Our updated prompt identifies six new scam types, shown in Table~\ref{tab:scam_types}. 

\vspace{.1cm}\noindent\textbf{Takeaway from \ref{rq1} and \ref{rq2}.}
%\marie{where does the conversation about scams come into the first sentence? please make explicit}
%Table~\ref{tab:suspicious_sms_distribution} point out that 
%Users primarily report suspicious text messages, followed by spam (Table~\ref{tab:suspicious_sms_distribution}).
%The categorization of all scam text messages shows that the
We identify fraudulent, non-fraudulent, and potentially malicious texts from user reports.
Over 40\% of reported texts are scams, with Wrong Number scams as the most reported type. This indicates that scammers are able to evade \mnonospace's XDR system using conversational scams instead of URL-based scams.
Alternatively, \mnos are better at blocking scam texts with URLs than ones without URLs. 
%We identify 40.27\% unique reports as scam texts and 35.12\% as spam from user-submitted reports. 
%This indicates that these scam text messages evade the \mnonospace's firewall detection the most. 
%Delivery/Parcel and Banking are the top scams for URL-based scams, whereas, for non-URL-based scams, Wrong Number and Banking scams are the two major categories. Scammers lure victims into scam texts with URLs differently compared to non-URL-based scams, motivating us to study the various lure principles and the readability of text messages in \S\ref{subsec:text_analysis}.
%Unsurprisingly, `Delivery/Parcel,' `Banking, and `Debt' scams dominate the scam text messages with URLs. 
%In addition to the six known scam types~\cite{agarwal_imc24}, 
%We identify 6 new scam categories from user reports. 
Our findings could help \mnos update their XDR system to block new scams and save users from falling prey. %to them.
%researchers update their considerations when investigating SMS scams. 

%\guillermo{I find the comparison of results with/without URLs potentially interesting, but we may want to start by presenting the aggregated results, show the volume of the different type of scams overall and then draw the attention to relevant differences when comparing messages with/without URLs. Then conclude by discussing the implications -- phising URLs are easy to fit in the narrative of Delivering/Parcel messages... this is, lure principles are different when scaming (without URLs) than when phishing someone (with URLs), and explain that this finding prompted us to study these lure principles more in details in what follows. }

\subsection{Originating Sender ID Distribution}
\label{subsec:senderid}

Unlike phishing over email, SMS messages or calls only have sender ID and timestamp as metadata. As discussed in \S\ref{subsec:data collection}, we collect the sender IDs reported by users, either forwarded by users to 7726 or automatically via Google's one-click reporting system, and analyze them to answer \ref{rq3}.

\subsubsection{Sender IDs}

This subsection investigates the sender IDs reported by users and abused by scammers to conduct scams. 

\vspace{.1cm}\noindent\textbf{Distribution.}
There are four types of sender IDs used to send SMS --- phone numbers, alphanumeric shortcodes, number-only shortcodes, and email addresses. 
Table~\ref{tab:sender_id_distribution} shows the sender ID distribution of text message reports. Unsurprisingly, the majority of these are phone numbers. Curiously, while 83\% of overall reports are from phone numbers, about 92\% of scams originate from phone numbers. This material discrepancy between reports and scam texts highlights the utility of dividing out scams and considering them separately. 

\begin{table}[!ht]
    \centering
    \scalebox{0.9}{
    %\resizebox{\columnwidth}{!}{%
\begin{tabular}{lrr}
    \hline
    \multicolumn{1}{c}{\bf Type} & \multicolumn{2}{c}{\bf Unique Sender IDs} \\ 
    & \multicolumn{1}{c}{\bf User Reports} & \multicolumn{1}{c}{\bf Scam Texts} \\
    \hline
    \hline
    Phone numbers & 103,301 & 79,894 \\
    Alphanumeric shortcodes & 19,318 & 6,047 \\
    Email addresses & 1,191 & 943 \\
    Number shortcodes & 429 & 207 \\
    \hline
\end{tabular}
%}
    }
    \caption{Distribution of all unique sender IDs (124,239 total) for all text message user reports and identified scam texts.}
    \label{tab:sender_id_distribution}
\end{table}

\begin{figure}[!ht]
    \centering
    \includegraphics[width=\linewidth]{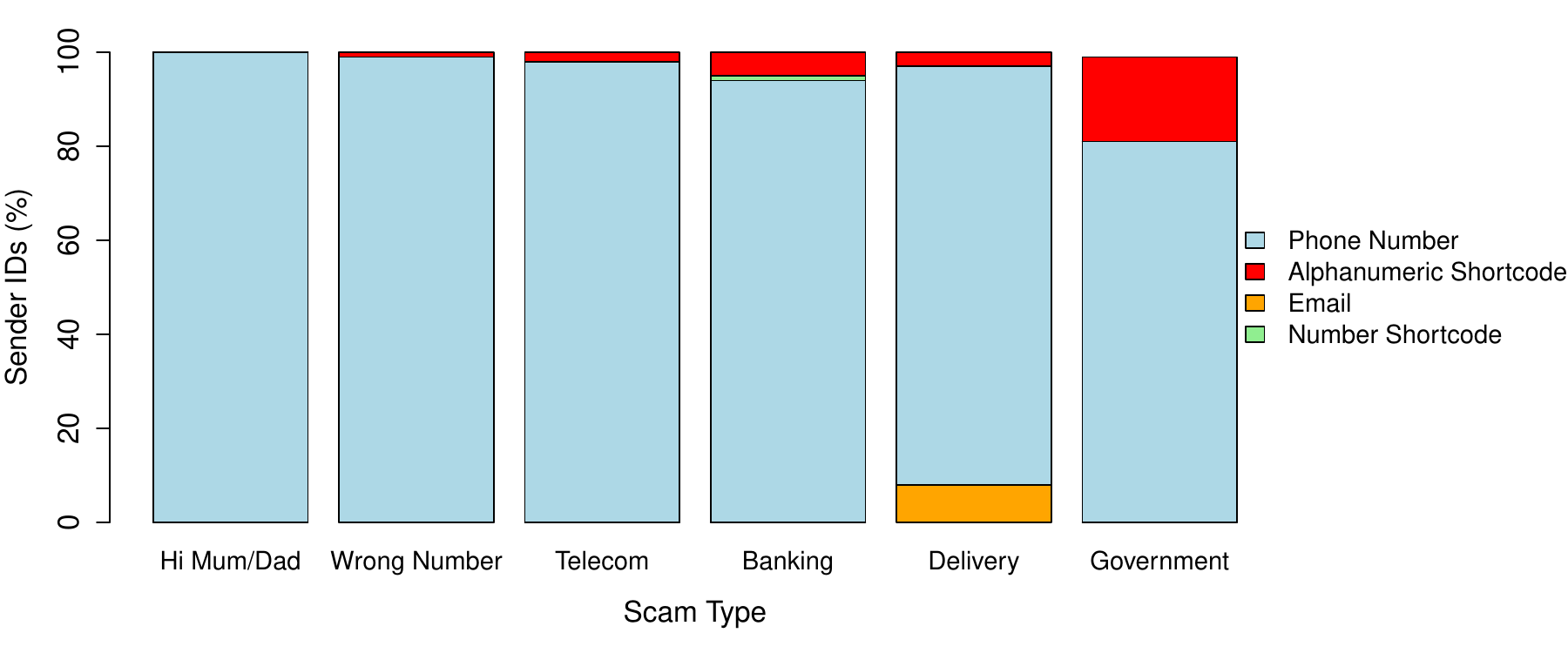}
    \caption{Sender IDs which scammers abuse to send six types of scam texts.
    %`Hi Mum/Dad' scams abuse only phone numbers, while email addresses are mostly abused to send Delivery/Parcel scams and alphanumeric shortcodes for government scams.
    Y-axis is normalized by the total number of scams in each category.} 
    \label{fig:sender_id_dis_scam_type}
\end{figure}

We analyze the distribution of sender IDs in relation to the types of scams to identify patterns that enhance our understanding of existing threats. We find that phone numbers dominate the breakdown by scam type. We notice that 18,179 phone numbers are abused to send Wrong Number scams, followed by 10,429 phone numbers for Delivery scams and 8,741 for Hi Mum/Dad scams. Due to the absence of Know-your-customer (KYC) checks in the UK, scammers can procure multiple SIM cards.

We also see cases where other sender IDs are abused. 
Fig.~\ref{fig:sender_id_dis_scam_type} shows that scammers abuse 875 email addresses to send delivery/parcel scams. The use of email addresses indicates that these scam text messages are sent to users on iMessage. The email address in the sender ID is available on Android devices but is more commonly used to send iMessage on Apple devices. Future work integrating one-click reporting on iMessage will show us if this trend holds with additional data.

While Delivery scams contribute toward the abuse of 363 alphanumeric shortcodes, 341 alphanumeric shortcodes are abused for banking scams. Scammers use similar-looking alphanumeric shortcodes to impersonate various delivery and banking entities to lure victims into the fraud: `evroi' instead of `evri' and `santandar' instead of `santander.'

\begin{figure}[!ht]
    \centering
    \includegraphics[width=0.8\linewidth]{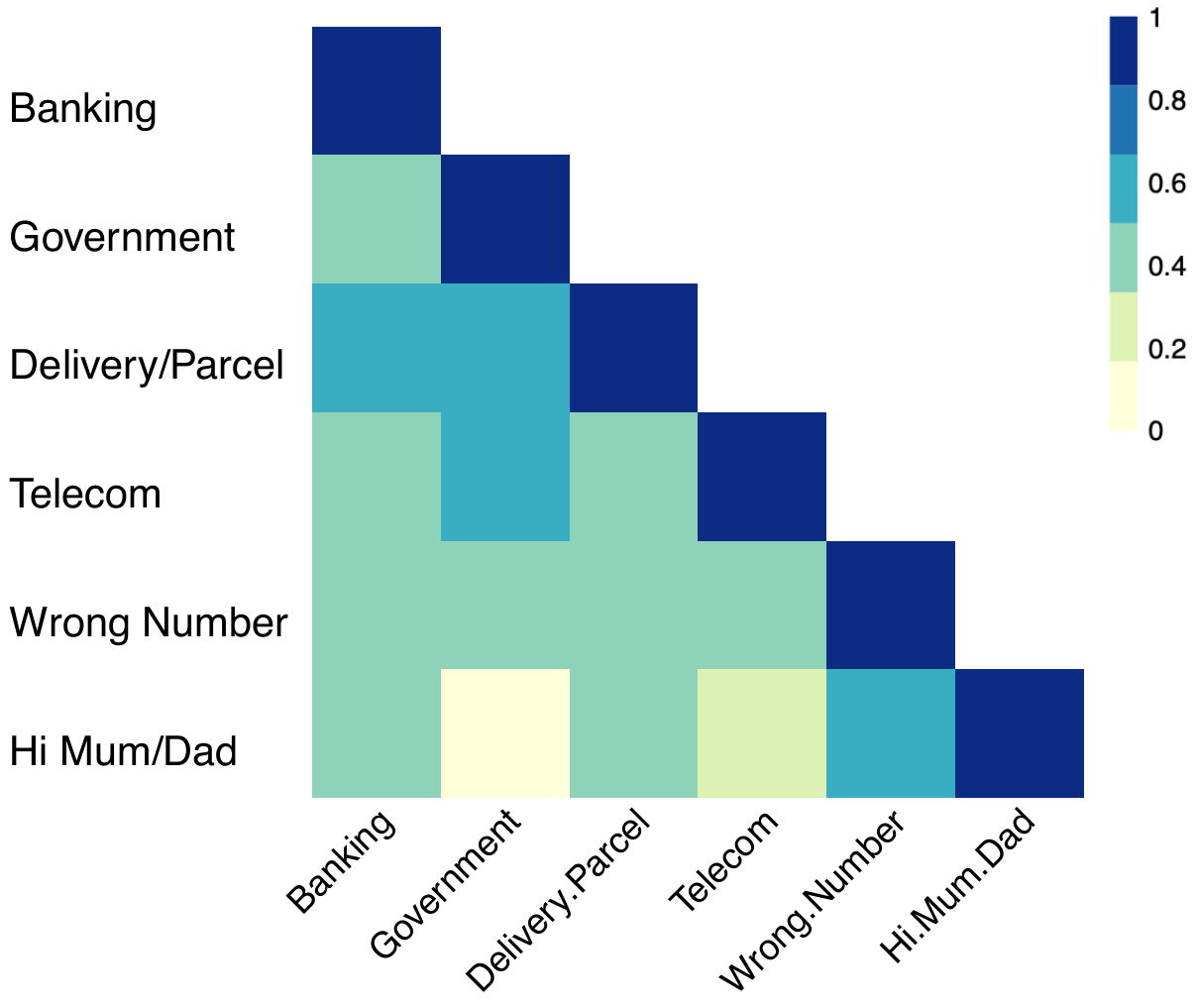}
    \caption{Heatmap of common Sender IDs used in different scams. Values normalized by the total common sender IDs.}
    \label{fig:sender_id_heatmap}
\end{figure}

We investigate overlapping sender IDs scammers abuse to send different scam types (Fig~\ref{fig:sender_id_heatmap}). We find 490 phone numbers abused to send both Hi Mum/Dad and Wrong Number scams. This indicates that some scammers conduct both types of conversational scams. 
An additional 183 sender IDs were used to send both Delivery and Banking scams (cf. Table~\ref{tab:common_senderids} in Appendix for all numbers). 
%This could be because the
One explanation could be scammers using the same third-party service to broadcast their scams.

\vspace{.1cm}\noindent\textbf{Countries.}
%As explained in \S\ref{subsec:data enrichment}, we perform a Home Location Register (HLR) lookup on all collected phone numbers. %HLR lookup validates the queried phone number and provides the country to which the phone number belongs. 
%To this end, we provide the distribution of the top 10 origin countries of the sender ID phone numbers from all user reports and the ones scammers abuse to conduct scams in Table~\ref{tab:origin_network_country_distribution}. We see 
%To this end, 
We analyze the origin of each phone number using our collected HLR lookup data.
We find that 73,815 (92.4\%) phone numbers abused to send scam texts originate in the UK.
%have the United Kingdom as the origin country. % significant proportion of the phone numbers originated in the United Kingdom.
This makes sense as it is easy and cheap to procure UK pay-as-you-go SIM cards without KYC. %Additionally, using the same country phone number is more believable when targeting users in the UK.
%, compared to international phone numbers. 
Additionally, it is more believable to target victims using the same country's number as theirs.
%In 2022, Ofcom issued a policy statement for UK \mnos to block international incoming scams~\cite{ofcom_guidelines,bbc_internationalblocking} %which helps explain the fewer international phone numbers sending scam texts.
%\marie{how is this relevant? Directly tie the ofcom regulation to either the current state or where it's going after it's under force.}

% \begin{table}[!ht]
%     \centering
%     \input{tables/origin_network_country_distribution}
%     \caption{Distribution of top 10 origin countries of the sender ID phone numbers from all text message user reports and identified scam text messages. \guillermo{I'd remove this table and just put the 95,860/73,815 figure in the text. Then you focus the discussion of this section aroud Table 7.}}
%     \label{tab:origin_network_country_distribution}
% \end{table}

%Investigating the mobile numbers abused to send different scam text messages,
Table~\ref{tab:scamtype_originnetwork_country} shows that more than 99\% phone number scammers abuse for the Wrong Number, Hi Mum/Dad, and Government scams originate in the UK
While more than 91\% of phone numbers originate from the UK for Delivery and Banking scams, we notice phone numbers originating from the USA and the Philippines abused for Delivery scams and numbers from Japan abused for Banking scams. This is likely due to preferences of a few attackers.
Scam campaigns impersonating only two particular UK banks account for almost all of the Japanense number (8.49\% of the total banking scams).
Similarly, we identify a campaign impersonating one Delivery brand with nearly identical text %that attributes to %3.09\% of phone numbers 
explaining most of the US numbers (and 3.09\% of delivery scams) and two campaigns originating from the Philippines, one of which belongs to the `darcula' iMessage and RCS smishing attacks~\cite{darcula_netcraft}.

% \guillermo{You need to wrap this paragraph by discussing reasons behind the differences in Table 7. Why Delivery/Parcel and Banking have more diverse originating countries? Can you swiftly review the 8\% and 3\% that originate in Japan and USA for Banking and Delivery/Parcel and offer a reason if there is something interesting to report? You may want to wrap up this discussion with the implications of this, small, yet interesting change in distribution.}

\begin{table*}[!ht]
    \centering
    \scalebox{0.9}{
    \resizebox{\linewidth}{!}{%
\begin{tabular}{l|lrlrlrlrlrlr}
    \hline
     \multicolumn{1}{c|}{\bf Scam Type} & \multicolumn{1}{c}{\bf Country} & \multicolumn{1}{c}{\bf \%} & \multicolumn{1}{c}{\bf Country} & \multicolumn{1}{c}{\bf \%} & \multicolumn{1}{c}{\bf Country} & \multicolumn{1}{c}{\bf \%} & \multicolumn{1}{c}{\bf Country} & \multicolumn{1}{c}{\bf \%} & \multicolumn{1}{c}{\bf Country} & \multicolumn{1}{c}{\bf \%}\\
    \hline
    \hline
    Wrong Number & UK & 99.14 & USA & 0.32 & Nigeria &0.09 & Canada & 0.06 & Ireland & 0.06\\
    Hi Mum/Dad & UK & 99.88 & Canada & 0.05 & Japan & 0.05 & Nigeria & 0.02\\
    Delivery/Parcel & UK & 91.72 & USA & 3.09 & Philippines & 2.85 & Thailand & 0.73 & Japan &0.33\\
    Telecom & UK & 95.12 & Philippines & 1.53 & Tajikistan & 0.98 & India & 0.68 & Vietnam & 0.3\\
    Banking & UK & 91.35 & Japan & 8.49 & USA &0.05 & Channel Islands & 0.04 & Indonesia & 0.02\\
    Government & UK & 99.37 & Japan & 0.21 & France &0.1 & Jersey & 0.1 & Poland & 0.1\\
    \hline
\end{tabular}
}
    }
    \caption{Distribution of top 5 origin countries of the sender ID phone numbers for six known scam types.}
    \label{tab:scamtype_originnetwork_country}
\end{table*}

\begin{figure}[!ht]
    \centering
    \includegraphics[width=0.8\linewidth]{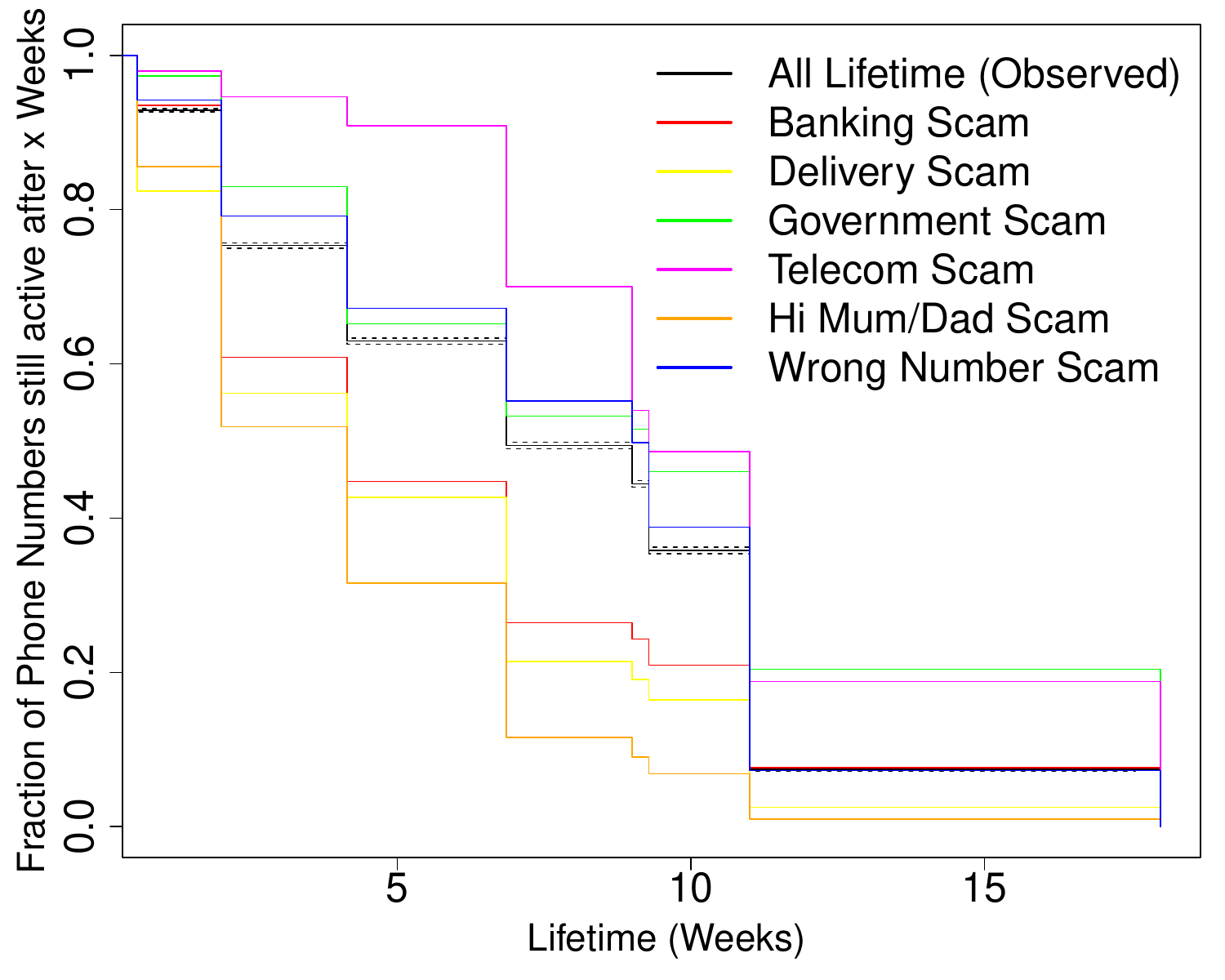}
    \caption{The number of weeks a mobile number (originating sender ID) is alive after being used to send a scam text message ($n=78,906$). Black dashed line is the 95\% CI for all phone numbers (observed) lifetime.
    %used to send identified scam text messages ($n=78,906$). The black dashed line represents the 95\% CI for all phone numbers (observed) lifetime.
    }
    \label{fig:survival_analysis}
\end{figure}

\vspace{.1cm}\noindent\textbf{Mobile Number Lifetime.}
%We receive weekly data from our collaborating \mnonospace. %As we run the HLR lookup on the mobile numbers weekly, we calculate the lifetime of the mobile numbers weekly. 
We study the lifetime of mobile numbers to understand how long scammers have used the same mobile number to scam their victims. 
To this end, we plot a survival curve to visualize the lifetime of the mobile numbers with overall survival probability in Fig.~\ref{fig:survival_analysis}. 
%The dotted lines in the plot are the 95\% confidence interval. 

We find that the median lifetime of all phone numbers scammers abuse is 48 days (6.86 weeks). 
While 75.3\% are active after 2 weeks, %While 62.9\% are active after a month, 
only 35.8\% are active after 10 weeks. 
For Hi Mum/Dad scams, we find the median to be 4.14 weeks, i.e., 29 days, whereas previous research found the median lifetime %of originating sender ID phone numbers for `Hi Mum/Dad' scams 
to be 14 days~\cite{agarwal_usenix25}. 
We attribute this difference to the fact that previous work only studies scam messages detected by existing XDR filters, missing those that permeate through their detection systems. 
Instead, our work investigates phone numbers reported by users. 
While we discover that the median lifetime of phone numbers abused to conduct delivery and banking scams is the same as the Hi Mum/Dad scams, the median lifetime for telecom is 9 weeks and over 9 weeks for wrong number scams. 
This indicates scammers being able to evade \mnonospace's XDR systems. %and continue operating their mobile numbers for a longer time. 
%Thomas et al. demonstrated that the median lifetime of phone numbers tied to abuse is less than one hour~\cite{kurt_voipabuse}. 
%While phone numbers investigated by Agarwal et al. for `Hi Mum/Dad' scams belong to scam texts that were already blocked by their partner \mnonospace, the phone numbers we examine are scam texts that evade our partner's firewall and reach the potential victims. 

\subsubsection{Mobile Network Operators}

%Home Location Register (HLR) lookup allows us to identify the \mno of a mobile number and its current status - (live/dead). 
\Mnos (MNOs) are one of the main stakeholders that scammers abuse to send scam texts and we use our HLR lookup data to uncover the \mno of a mobile number.
Table~\ref{tab:mno_distribution} presents the top 10 different \mnosnospace.\footnote{We refrain from naming \mnos in line with the confidential agreement with our partner.} 
\Mnos can be (1) Physical --- issues a physical SIM card, (2) Virtual --- does not issue a physical SIM card, and (3) Mobile Virtual Network Operator (MVNO) --- issues a physical SIM card but 
%does not own their own mobile network, rather 
rents space from other \mnos instead of running their own.

UK \mnos that provide physical SIM cards are more abused than Virtual \mnosnospace. This could be because scammers broadcast thousands of scam texts using a SIM box/farm~\cite{dcpcu_simfarm}. A SIM box is an SMS gateway device connected to a computer with multiple SIM slots. MNO 1 is the most preferred \mnonospace, followed by MNO 2, 3, and 4. While MNO 1, 2, and 3 support GSM technology that works with SIM boxes, MNO 4 does not support GSM.
%Other factors that could influence the popularity of \mnos abused by scammers 
The desirability of \mnos by scammers could also depend on the ease of availability to procure the SIM cards, network availability where the scammers are based, or the XDRs that scammers evade to detect these scam texts.

\begin{table}[!ht]
    \centering
    \scalebox{0.9}{
    \resizebox{\columnwidth}{!}{%
\begin{tabular}{lll|r|r}
    \hline
    \multicolumn{1}{c}{\bf MNO} & \multicolumn{1}{c}{\bf Country} & \multicolumn{1}{c|}{\bf Type} & \multicolumn{1}{c|}{\bf Current} & \multicolumn{1}{c}{\bf Original}\\
    &&&\multicolumn{1}{c|}{\bf MNO} & \multicolumn{1}{c}{\bf MNO} \\
    \hline
    \hline
    MNO 1 & UK & Physical & 19,576 & 25,629\\ %O2 UK
    MNO 2 & UK & Physical & 13,572 & 20,938\\ %EE/BT
    MNO 3 & UK & Physical & 9,885 & 12,051\\ %vodafone uk
    MNO 4 & UK & Physical & 8,718 & 8,614\\ %Three
    MNO 5 & UK & Virtual & 1,318 & 1,318\\ %Tismi BV
    MNO 6 & UK & Physical (MVNO) & 1,233 & 769\\ %Sky 
    MNO 7 & UK & Physical (MVNO) & 999 & 2,136\\%Lyca
    MNO 8 & UK & Virtual & 619 & 654\\%Gamma
    MNO 9 & UK & Virtual & 296 & 289\\%Stour Marine
    MNO 10 & USA & Physical & 292 & 134\\%T-Mobile (US)
    % original network: 
    % Numbers Telecom Limited - 4212
    % Voxbone SA - 545 - landline
    % Citrus - 514
    % Ziron - 434
    % AQL - 399 - landline
    % Colt - 363 - landline
    \hline
\end{tabular}
}
    }
    \caption{Top 10 \mnos abused to send scam text messages. {\em Current MNO} is the network where the phone number is currently assigned. {\em Original MNO} is the network assigned to this telephone number range.}
    \label{tab:mno_distribution}
\end{table}

Next, we explore the popularity of \mnos based on scam types. While MNO 1 is least abused when sending Telecom scams, MNO 2, 3, and 4 are least abused when it comes to Hi Mum/Dad and Government scams. This could mean that MNO 1 is better at detecting Telecom scams, whereas MNO 2, 3, and 4 are better at blocking Hi Mum/Dad and Government scams.  
While most Virtual numbers are primarily abused to send Wrong Number scams, MVNOs are abused for Telecom, Hi Mum/Dad, and Wrong Number scams. Virtual numbers provide scammers the advantage of sending scam texts from outside the UK. However, with MVNOs, even though they use the network of a physical MNO, the text messages go through their own XDR systems (if implemented). 
Looking into MNO 10, the only non-UK \mno in Table~\ref{tab:mno_distribution}, we find similar text messages targeting one Delivery company. This indicates that a single threat actor abused 3.09\% of all sender IDs that belong to the USA.
% MNO 10 is the only non-UK \mno in the top 10 list of \mnos scammers abuse. Table~\ref{tab:scamtype_originnetwork_country} demonstrates that in Delivery scams, 3.09\% sender IDs belong to the USA, attributing to MNO 10.
For Hi Mum/Dad scams, MNO 1 is significantly more abused (86.4\%) than others~\cite{agarwal_usenix25} showcasing that scammers abuse different types of \mnos for various scams and some more than others.
%\marie{Idk how the USA fits into this. Please add more context. Feels a bit random.}

\vspace{.1cm}\noindent\textbf{Takeaway from \ref{rq3}.}
%Scammers primarily abuse phone numbers to send scams. 
We find that scammers prefer certain MNOs over others depending on scam types. This indicates that MNO's XDR systems use independent rules and do not share intelligence. 
We suggest MNOs work collaboratively and share their best practices and XDR rules to make the blocking a collective effort. %Most phone numbers are abused by Wrong Number scams, followed by Delivery and Hi Mum/Dad scams. 
%While email addresses are primarily abused when sending delivery scams, alphanumeric shortcodes are abused mostly by delivery and banking scams. 
Investigating user reports can help MNOs identify and block new originating sender IDs so scammers cannot abuse them for long. 
As scammers also abuse shortcodes, we suggest implementing a central sender ID registry operated by the regulator that could stop the abuse of impersonation scams.

\subsection{Domain Analysis}
\label{subsec:domain_analysis}

Scammers trick victims into clicking on malicious URLs sent in text messages and lure them into providing their personal or financial information. 
We examine the various infrastructure services scammers abuse to host phishing pages shared via scam texts, answering \ref{rq4}.

\vspace{.1cm}\noindent\textbf{URL Shorteners.}
We identify a significant amount of URLs in the text message reports that are shortened URLs, inline with prior research~\cite{nahapetyan2024sms}. Table~\ref{tab:url_shorteners} presents the number of unique URLs belonging to the 10 most popular URL shorteners abused to conduct scams, with the two most common services being \url{bit.ly} and \url{t.ly}.
Scammers abuse URL shortening services to evade detection from XDRs and antivirus vendors.
%S%cammers set up shortened URLs as they can be 
Shortened URLs also can be created for free, fit in an SMS character limit, and redirect to the malicious URL only when a user clicks. 

\begin{table}[!ht]
    \centering
    \scalebox{0.9}{
    \resizebox{\columnwidth}{!}{%
\begin{tabular}{l|r|rrrrrr}
    \hline
    \multicolumn{1}{c|}{\bf URL} & \multicolumn{1}{c|}{\bf Unique} & \multicolumn{6}{c}{\bf Scam Types} \\%& \\
    \multicolumn{1}{c|}{\bf Shortener} & \multicolumn{1}{c|}{\bf URLs} & \multicolumn{1}{c}{\bf W} & \multicolumn{1}{c}{\bf H} & \multicolumn{1}{c}{\bf T} & \multicolumn{1}{c}{\bf B} & \multicolumn{1}{c}{\bf G} & 
    \multicolumn{1}{c}{\bf D} \\%& \multicolumn{1}{c}{\bf Scam Acronyms} \\
    \hline
    \hline
    bit.ly & 2,280 & 0 & 0 & 12 & 50 & 48 & 73 \\%& \multicolumn{1}{l}{W - Wrong Number}\\
    tinyurl.com & 1,094 & 0 & 0 & 48 & 8 & 16 & 241 \\%& \multicolumn{1}{l}{Hi - Hi Mum/Dad}\\
    is.gd & 814 & 0 & 0 & 5 & 1 & 2 & 393 \\%& \multicolumn{1}{l}{T - Telecom}\\
    wa.me & 712 & 8 & 190 & 1 & 4 & 3 & 0 \\%& \multicolumn{1}{l}{B - Banking}\\
    rb.gy & 611 & 0 & 0 & 185 & 2 & 6 & 250 \\%& \multicolumn{1}{l}{G - Government}\\
    cutt.ly & 462 & 0 & 0 & 95 & 4 & 13 & 61 \\%& \multicolumn{1}{l}{D - Delivery}\\
    qrco.de & 447 & 0 & 0 & 82 & 1 & 3 & 354\\
    rebrand.ly & 434 & 0 & 0 & 314 & 2 & 1 & 21\\
    t.ly & 302 & 0 & 0 & 5 & 16 & 0 & 49\\
    tiny.cc & 18 & 0 & 0 & 0 & 0 & 2 & 0\\
    \hline
\end{tabular}
}
    }
    \caption{Distribution of Top 10 URL shorteners abused by scammers to send scam texts. (W: Wrong Number, H: Hi Mum, T: Telecom, B: Banking, G: Government, D: Delivery)}%\guillermo{Move acronyms to the caption and try to fit this table in one column}}
    \label{tab:url_shorteners}
\end{table}

In addition to the URL shorteners in Table~\ref{tab:url_shorteners}, we identify 173 unique `.sbs,' and 62 unique `.me' URLs abused in scam text messages. For example, we find six second-level domains that try to impersonate EVRI that contain `evri' plus one more character as the second-level domain name for `.sbs' URLs.
The registrar cannot take down malicious shortened URLs; deleting the entire domain would cause harm to other, non-malicious shortened URLs.
Instead, a URL shortener %being a third-party service
requires investigating and take down on URL shortener service's side. 

\vspace{.1cm}\noindent\textbf{Registrars.}
We find that NameSilo (26.3\%) is the most abused registrar for SMS scam URLs, followed by Hosting Concepts B.V. (13.2\%). 
Previous research on newly registered phishing domains also found NameSilo as the most abused registrar~\cite{agarwal25:weis}. On the other hand, a smishing research using US smish reports identified NameCheap~\cite {smishtank_timko24} as the most abused registrar. %s for similar types of fraud.
This highlights differences in our collected data as well as the impact of our spam/scam distinction.
%This indicates that certain registrars are comparatively more lenient towards scammers abusing their services to register new domains.
While domains in delivery and telecom scams primarily abuse NameSilo, MarkMonitor is the most abused registrar for banking and government scams. This highlights that different scams abuse different registrars, likely reflecting preferences of different groups of threat actors.

\vspace{.1cm}\noindent\textbf{Top-level Domains (TLDs).}
We present the ten most abused top-level domains (TLDs) by scammers to register domains abused in scam text messages in Table~\ref{tab:tld_scams}. We differentiate TLDs 
by domains abused by scammers and domains with URL shorteners as URL shorteners are third-party services, so their TLDs should not misunderstood for abuse. %, similar to domains that scammers abuse. 

We find that `.com' remains the most abused TLD, followed by `.top' and `.co.uk.' This is in line with recent research~\cite{agarwal25:weis, korczynski2017reputation, domainz_sp16, kumar2021diversity}. 
Investigating individually by scam types -- `.com' remains the most abused for delivery, banking, telecom, government, and wrong number scams. After `.com,' we uncover that the delivery scams abuse .top followed by .xyz. On the other hand, we find that government scams abuse `co.uk' and `.uk' TLDs making them more convincing for potential victims. For example, \url{arrange-test-kit[.]co[.]uk} is used to impersonate a health service text. While `.buzz' is the second most abused TLD by scammers for telecom scams, they abuse `web.app' for banking scams. For example, scammers set up \url{attempted-logon[.]web[.]app} to lure victims into clicking on the malicious link.

\begin{table}[!ht]
    \centering
    %\resizebox{\columnwidth}{!}{%
\begin{tabular}{lr|lr}
    \hline
    \multicolumn{2}{c|}{\bf URLs} & \multicolumn{2}{c}{\bf URL Shorteners}\\
    \multicolumn{1}{c}{\bf TLD} & \multicolumn{1}{c|}{\bf Unique URLs} & \multicolumn{1}{c}{\bf TLD} & \multicolumn{1}{c}{\bf Unique URLs}\\
    \hline
    \hline
    com & 3,593 & ly & 3,480\\
    top & 897 & com & 1,097\\
    co.uk & 390 & gy & 611\\
    xyz & 370 & gd & 814\\ 
    buzz & 166 & me & 712\\
    info & 156 & de & 447\\
    web.app & 143 & cc & 18\\
    sbs & 140 & co & 15\\
    cyou & 101 & ws & 7\\
    \hline
\end{tabular}
%}
    \caption{Top 10 Top-level Domains (TLDs) abused in all unique scam URLs.}
    \label{tab:tld_scams}
\end{table}

\vspace{.1cm}\noindent\textbf{Autonomous Systems (ASes).}
We identify 56,092 unique IP addresses that the identified scam domains abused. We find that 7,110 (12.7\%) unique IP addresses belong to Cloudflare, a proxy service which hides the server IP address. Previous research also found most domains in their dataset abusing Cloudflare~\cite{nahapetyan2024sms, agarwal25:weis}. 
Table~\ref{tab:as_geographical_distribution} presents the five most abused Autonomous Systems (ASes) to host the identified scam URLs, excluding Cloudflare. It also shows the geographical location of the IP addresses abused to host scam domains.

While scam messages target individuals in the UK and significantly dominate UK phone numbers used to send scam texts, surprisingly, we only see 1,267 IP addresses abused to host domains based in the UK. It might be that scammers abuse the infrastructure outside their target country to make it challenging for takedown companies and law enforcement, giving themselves more time before the domain is taken down. 
%Although Amazon is the largest abused AS, surprisingly 51.6\% IP addresses are based outside the US.  
%\marie{I don't connect these dots. Could you explain a bit more?}

\begin{table}[!ht]
    \centering
    %\resizebox{\columnwidth}{!}{%
\begin{tabular}{lr|lr}
    \hline
    \multicolumn{1}{c}{\bf AS} & \multicolumn{1}{c|}{\bf Unique IPs} & \multicolumn{1}{c}{\bf Country} & \multicolumn{1}{c}{\bf Unique IPs}\\
    \hline
    \hline
    Amazon & 42,266 & United States & 20,452\\
    Akamai & 1,824 & China & 9,741\\
    Hostinger & 1,463 & India & 8,080\\
    Cogent & 502 & Brazil & 4,069\\
    Tencent & 166 & Ireland & 2,801\\
    % Alibaba & 83 & Netherlands & \\
    % NTT America & 78 & France & \\
    % Namecheap & 59 & Japan &\\
    % Google & 52 & & & \\
    \hline
\end{tabular}
%}
    \caption{Top 5 Autonomous Systems (ASes) and countries where scam URLs are hosted, excluding Cloudflare.}
    \label{tab:as_geographical_distribution}
\end{table}

% \begin{table}[!ht]
%     \centering
%     \input{tables/ip_country}
%     \caption{Top 10 countries where the identified IP addresses abused to host unique scam URLs are based, excluding Cloudflare ($n=58,657$).}
%     \label{tab:ip_country}
% \end{table}

\vspace{.1cm}\noindent\textbf{Domain Randomness.}
We evaluate the randomness of second-level domains (SLDs) abused in smishing attacks.
We find that 60.3\% SLDs contain a (nontrivial) dictionary word, indicating use of meaningful words.
This may imply the use of words such as `cancel,' `track,' and `ship' along with brand names (including typosquatted brand names), which would be consistent with common impersonation tactics inline with previous research~\cite{agten2015seven, agarwal25:weis, moore2010measuring, szurdi2014long, kintis2017hiding}.
While the mean Shannon entropy is 2.88, suggesting moderate randomness, the average vowel-to-consonant ratio is 0.64, higher than expected for randomly generated strings. This indicates a bias toward pronounceable patterns. 
We select a random sample of 100 domain names and find 67\% to be brand typos (e.g., evrlgb-couriers, verify-myapplepay, hsbc-cancel-payment, icloud-uk), also known as combosquatting~\cite{kintis2017hiding}.
%On the contrary, we identify 33\%
The rest contain arbitrary character sequences (e.g., dcfcy, onapuw, azirew, acozir).

\begin{figure}[!ht]
    \centering
    \includegraphics[width=.8\linewidth]{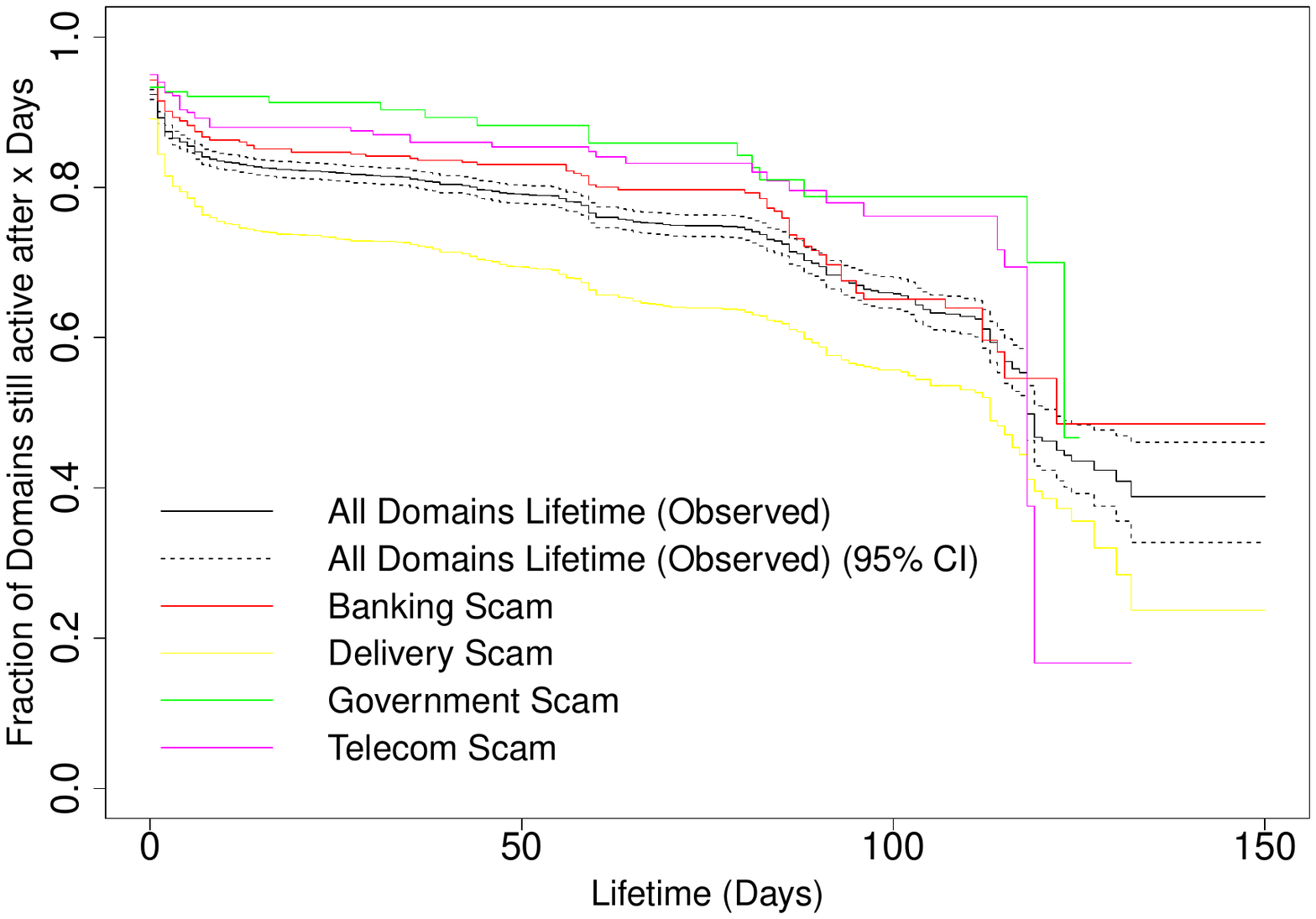} 
    \caption{Survival analysis of the lifetime of domains scammers abuse to conduct SMS scams.}
    \label{fig:ecdf_scam_domains}
\end{figure}

\vspace{.1cm}\noindent\textbf{Domain Lifetime.}
We find the lifetime of 5,854 unique domains (excluding URL shorteners), with a median of 118 days. 
We use survival analysis to investigate lifetime of all identified scam domains with overall survival probability (Fig.~\ref{fig:ecdf_scam_domains}). The dotted lines in the plot are the 95\% confidence interval.
%ranging between 0 and 3,728 days, with a median of 59 days. To understand the number of days the identified domains are active, we analyze the distribution of their lifetime as seen in .%, with the X-axis on a log scale.
While 83.3\% of the domains are active after 10 days, 65.8\% are active after 100 days. This demonstrates that longer lived domains live longer.
Previous work found that more than 40\% of the domains from the US Federal Trade Commission (FTC) user reports were active for over 100 days~\cite{srinivasan2016understanding}. On the contrary, Nahapetyan et al. observe an average lifetime of 12.241 days with a median of 0.53 hours~\cite{nahapetyan2024sms}. Others identify the average lifetime of scam domains as 59.4 days~\cite{kotzias2025ctrl+}. 

The aggregate lifetime of all domains represents a broad view; we look into the domains scammers abuse to conduct different scams.
Delivery scam domains have a median lifetime of 113 days, while government scam domains last longer, with a median of 123 days.
For delivery scams, 55.7\% of domains remain active after 100 days, but 78.7\% of domains abused to conduct government scams are active after 100 days.
For example, \url{tax-rebate[.]top} was active only for 9 days, while \url{online-uk-claims[.]com} remained active for 123 days. 
This shows the different takedown practices or domain registration preferences for different types of scams.

Particularly, the identified domains last so long as scammers register and re-register domain names, continuing to abuse them over time~\cite{hao_imc13}. For example, we see \url{uk-delivery[.]com} was first seen on April 20, 2014, and has been likely re-registered multiple times since then.
%Moore and Clayton discovered that scammers re-register bank domains after banks close, taking advantage of the dead bank's reputation for illicit activities~\cite{moore2014ghosts}. 

\vspace{.1cm}\noindent\textbf{Takeaway from \ref{rq4}.}
 %, such as Cloudflare. 
Scammers abuse third party services like URL shorteners and proxy services to evade detection.
Infrastructure services should perform Know-Your-Customer (KYC) checks before allowing users to use their services and collaborate with threat intelligence organizations and \mnos to perform more effective takedowns. 
Domains abused to conduct scams have a median lifetime of 118 days, indicating that scammers are able to keep the domains active for a long time.
Registrars need to proactively identify scam domains, particularly potential re-registeried scam domains, and perform takedowns on abused domains.

%domains that have been previously detected for scams%, like \url{online-uk-claims[.]com}, first seen on Dec 14, 2020.
%It also highlights the need to differentiate between compromised and new (re)registered domains as the takedown strategies differ.

\subsection{SMS Text Analysis}
\label{subsec:text_analysis}

The text of the scam message provides insights into how scammers craft these messages to deceive victims into taking action. This subsection answers \ref{rq5} by working to understand the various lures and readability of the scam messages.

\vspace{.1cm}\noindent\textbf{Scam Lures.}
Scammers design text messages to deceive victims, asking them to take action. 
%We work to understand this deception using a framework from social science developed by Stajano and Wilson.
%As explained in \S\ref{subsec:textclass}, 
We work to understand this deception by classifying the scams by lure using the typology identified by Stajano and Wilson~\cite{StajanoFrank2011_principles}. %~Fig.~\ref{fig:ecdf_scam_domains} s.
Table~\ref{tab:lure_principles} %describes our typology and
presents the scams we classify by lures. Scammers use the time/urgency lure in all scam types, forcing users to make an impulsive decision. For example, using words like `now,' `urgent,' or `immediately.'
We find that scammers lure victims into Hi Mum/Dad scams by also employing distraction and kindness, similar to prior findings~\cite{agarwal_usenix25}. For example, scammers pretend to be a victim's child, provide a random reason for reaching out from a different number and request immediate help. Similarly, for Wrong Number scams, scammers %use distraction and kindness principles, 
provide unrelated details and asking a random question or sounding like a friend to deceive potential victims into replying, which can turn into a pig butchering scam~\cite{ordekian:ecrime24}. 

\begin{table*}[!ht]
    \centering
    \resizebox{\linewidth}{!}{
\begin{tabular}{llr}
    \hline
    \multicolumn{1}{c}{\bf Scam Lure} & \multicolumn{1}{c}{\bf Definition} & \multicolumn{1}{c}{\bf Scam Types}\\
    \hline
    \hline 
     Authority & Scammers refer to trusted third parties and convince users to do things & Banking, Government,\\
     & they would not do otherwise & Delivery/Parcel \& Telecom\\
     Dishonesty & Scammers invite users willingly and knowingly into taking fraudulent action & - \\
     Distraction & Scammers provide unrelated details to distract the user & Hi Mum/Dad \& Wrong Number\\
     Need \& Greed & Scammers leverage users' greed and offer attractive benefits & Banking, Government \& Telecom\\
     Herd & Scammers convince that others have won taking the same risk& - \\
     Kindness & Scammers leverage the willingness of people to help others & Wrong Number \& Hi Mum/Dad\\
     Time \& Urgency & Scammers put time pressure on users so they make an irrational decision & All\\
    \hline
\end{tabular}
}
    \caption{Scam types categorized by lures (adapted from Stajano and Wilson~\cite{StajanoFrank2011_principles}).}
    \label{tab:lure_principles}
\end{table*}

% \guillermo{This section does not seem to bring anything new. Have the lure principles changed over time? Perhaps looking at temporal dynamics could bring something more substantial into this section?}

%Scammers also use the distraction principle in Hi Mum/Dad and Wrong Number scams by providing too many unrelated details and deceiving victims into replying. 

Scammers use need and greed to lure victims into Government (12.6\%), Telecom (13.7\%) and Banking (11.7\%) scams by offering tax refunds, points, cashback, or mentioning a suspicious payment, tempting the user to take the requested action in the text. 
Government (67\%), Telecom (47.1\%), Banking (44.2\%), and Delivery (37.8\%) scams also invoke authority, making it more convincing by impersonating government and private entities such as HMRC, EE, Barclays, or EVRI. This indicates that scammers do not just randomly draft scam texts but take advantage of various lures to deceive potential victims into taking an action. Depending on the type of scam, scammers use one or more lures in the text of the message. 
Unsurprisingly, we do not find SMS scammers using herd and dishonesty lures despite these commonly used in cryptocurrency scam ads~\cite{gilberto_ecrime22,agarwal2023_fc}.
%2278, 796, 518
%8636, 5506, 2755, 2745

\vspace{.1cm}\noindent\textbf{Gunning Fog Index.}
Scammers design the text of the scam messages depending on their target victims. The text message's readability is essential as not all potential victims can understand complex sentences in scam text messages. To target a broader range of individuals, one would expect the scam text messages to be simple. Towards this end, we use the Gunning Fog Index (GFI), designed to compute the years of education required to understand a given passage/text~\cite{gunning1969fog}. 
%Using the methodology explained in \S\ref{subsec:analysis methods}, 
We find that the mean of the Gunning Fog Index for conversational scam text messages is lower than others. While Hi Mum/Dad and Wrong number scams have a Gunning Fog Index of 4.6 and 3.9, respectively, Delivery/Parcel, Banking, Telecom, and Government impersonation scams have the Gunning Fog Index above 5.6, the maximum being 8.3 for Delivery/Parcel scams (Fig~\ref{fig:gfi_plot} in Appendix shows the GFI for all scam types). 
The Gunning Fog Index does not significantly differ between scam types. This indicates that scammers design straightforward scam texts that anyone with a maximum of 8th-grade education can read and understand.

% Delivery/Parcel - 7.8 (w/o URL), 8.3 (w/ URL)
% Banking - 6.2 (w/o URL), 5.6 (w/ URL)
% Telecom - 6.7 (w/o URL), 5.8 (w/ URL)
% Government - 8.2 (w/o URL), 7.1 (w/ URL)

%\sharad{Add a plot for the GFI}

\vspace{.1cm}\noindent\textbf{Takeaway from \ref{rq5}.}
%Scammers deceive victims into engaging in different scams based on various lure principles. %Understanding the lure principles scammers use for various scams can be used to educate and make users aware. 
%or example, 
Dividing scams by lures used can help educate victims on common ruses.
Scammers use authoritative lures for Banking, Government, Delivery, and Telecom scams. Users should ensure messages are from legitimate senders with trustworthy URLs. % with no grammatical errors in the organization/brand name mentioned. 
%kindness and distraction principles are primarily used for Wrong Number and Hi Mum/Dad scams. Unsurprisingly, scammers use the time/urgency principle for all scams but do not use the Herd or Dishonesty principle for any SMS scam.
%The Gunning Fog Index results show 
We find that most scam texts can be read and understood by those with a middle school education level, with no real differences between types.
%We find no significant differences in the education level required to understand types of scam texts.
%This indicates that the Index is probably not designed to study scam texts, as they are short text messages. 
%\marie{relate lure principles}

\section{Discussion}
\label{sec:discussion}

We next discuss findings we derive from analyzing user text reports in the context of our limitations. 

\vspace{.1cm}
{\noindent \bf Differentiating fraudulent from non-fraudulent messages is crucial in an increasingly challenging landscape}. 
Our analysis reveals that over 40\% of text messages are scams, observing on the other end a significant number of spam reports. 
However, we see that messages with fraudulent intent show different patterns than those without. 
Quite a number of our findings vary with intent like the likelihood of using a phone number vs. other sender IDs or the likelihood of using a URL. 
Previous research fails to differentiate between the scam and spam text messages, limiting its effectiveness in combating smishing~\cite{srinivasan2016understanding,tang:ccs22,smishtank_timko24,nahapetyan2024sms}. 
This distinction is of heightened importance when factors like the infrastructure or text patterns used are key ones that feed into XDRs at \mnosnospace. 
When these factors differ, % from the ones use to spread spam, %, as we see in our study. \guillermo{FUTURE: We are missing a comparative analysis with the infrastructure used in SCAM. After the deadline, consider running queries in ASs, HLR, etc for SPAM in case we need it for the rebuttal}
stakeholders require to adopt distinct strategies to address fraudulent messages effectively using user reports~\cite{spamshield_mavenir}. 
We call upon other researchers to also make this vital distinction in order to positively protect users from harm.

\vspace{.1cm}
{\noindent \bf Scammers shift toward non-URL-based lures, which bypass \mnosnospace's XDR systems}.
With the uptake in SMS scams, we see ``Wrong Number'' and other conversational scams dominating the submissions --- here we identify six new scam types above narrower prior work~\cite{agarwal_imc24,agarwal_usenix25}. 
These new types of conversational scams bypass \mnosnospace' XDR systems, which are more effective at detecting URL-based threats. 
This indicates a critical gap in current detection mechanisms, as most reported scams are texts without URLs, and shows the need to enhance algorithms for identifying such messages. 
This feature also highlights a key distinction from email phishing, posing challenges to the adoption of effective defense mechanisms~\cite{oset:ecrime18, sahingoz2019machine}. The shift in SMS scams echoes an evasion strategy similar to technical support scams, which have advanced from URL-based scareware URLs~\cite{stone2012underground} to call-based attacks via malvertising~\cite{miramirkhani2016dial}, bypassing conventional URL-focused defenses. 

We note that Telecom regulators maintain a list of phone numbers (aka do not originate or DNO) from organizations like banks to combat call spoofing~\cite{ofcom_dno}. 
We suggest that \mnos query phone numbers in the text of an SMS that posits to be an organization against the DNO list as an effective legitimacy checking mechanism, in the direction of Domain-based Message Authentication, Reporting \& Conformance (DMARC) used in mail servers~\cite{kucherawy2015domain} and telephony blocklists~\cite{pandit2018towards}.

\vspace{.1cm}
%{\noindent \bf Scammers abuse email addresses and phone numbers to send RCS/iMessage scams}.
{\noindent \bf Exploiting online encrypted messaging platforms to send scam texts}.
We also notice scammers use email addresses for delivery scams indicating RCS/iMessage scams. Rich communication service (RCS) and iMessage allow users to send encrypted SMS over the internet instead of SMS over \mnonospace. 
In addition to emails, we identify a scam campaign with over 44 deliver scam texts known to send RCS/iMessage scam texts called `darcula' --- a phishing-as-a-service platform~\cite{darcula_netcraft, akyazi2021measuring}. As \mnos improve their XDRs, scammers are turning to similar and cheaper alternatives like iMessage and RCS to bypass detection. 
The shift in smishing from traditional SMS to encrypted messaging platforms mirrors how traditional email-based phishing~\cite{moore2007empirical} has evolved towards cryptocurrency platforms~\cite{chendissecting, chen_cryptophish}. 
As RCS/iMessage texts directly bypass the \mnonospace's XDRs, we suggest services like Apple and Google collaborate with \mnos to identify and proactively block scams, enabling a more coordinated and comprehensive mitigation strategy.

\vspace{.2cm}
{\noindent \bf Limitations}.
We receive the user reports from only one %UK 
\mnonospace. 
%\guillermo{I liked the following paragraph (now hidden), but we need the space. Bring back if we manage to free up more}
%The research community needs access to data from all UK \mnos to provide an accurate distribution of all sender IDs and scams, but we cannot confirm that callback SMS scams and the DNO lists would not be available to other researchers out of the UK. 
%While there is no known reliable method to differentiate between RCS and SMS messages, utilizing the sender ID and 160-character text limit, we can guess 14\% of all unique reports to be potential RCS messages. 
%
The unavailability of %categorized
SMS scam data is a general limitation here. %in the area. 
We make significant progress on this limitation 
for one \mno by devising a processing pipeline that can characterize popular, known scam types and identify novel ones. 
As a key contribution, we see that the lures scammers use to deceive victims into taking an action differs based on the scam type.
Understanding the lures used could assist interveners to create more effective warnings and better user education about ongoing scams. Interveners include private companies like the Students Loans Company~\cite{student_smishing} and governments~\cite{ftc_howto}.
%The understanding of lures could assist organizations in creating effective warnings and better-educating users about ongoing scams. 
%For example, the Students Loans Company advises students at the start of the academic year about smishing scams~\cite{student_smishing}.

While we identify spam and scams, we also flag 23.26\% text reports as `Unknown.' 
Even though the URLs in the reports are not flagged by an AV vendor, previous research has shown that more than 94\% of the blacklisted domain names do not appear in public blacklists for several weeks or even months after they are first reported in abuse complaints~\cite{srinivasan2016understanding}. 
There is room for improving our text characterization and scam type identification. 
Some of the misclassifications we observe reveal fraudulent messages mimicking legitimate messages, with some being as simple as `Hi' --- requiring more sophisticated detection mechanisms. 

However, the scope of our paper is not to devise such detection mechanisms, but to offer a first look into SMS scams and scammer strategies through the lens of a large base of recent user reports. 
Our evaluation metrics (88\% for spam messages, 89\% for known scam types, 80\% new types) demonstrate high standards to answer our research questions, especially considering the scale and real-world conditions of our evaluation. 
While a deeper understanding of {\it unknowns} and {\it other} new types of scams is the scope of our future work, we reveal six emerging scam types (that group together) (§\ref{subsec:characterization_user_reports}); finer breakdowns will be a valuable future direction.
A challenge to address in this direction is to adopt novel detection mechanisms~\cite{pimentel2014review} to identify and classify new trends. 

%\vspace{.1cm}
%{\noindent \bf OpenAI is unable to classify 44.12\% scams that we mark as `Others'}.
%Investigating the text messages marked as `Others', we find OpenAI is unable to classify some simple text messages such as `<brand name> Please visit [URL].' On the other hand, we also see reported texts that should be classified as Wrong Number scams. While, we create our prompt based on known facts about Wrong Numbers scams, scammers create messages ranging from `Hi' to a sentence that would make a user reply. This indicates that we need novelty detection mechanism~\cite{pimentel2014review} to identify and classify new scam types. 

\vspace{0.2cm}
\noindent\fbox{%
    \centering
    \parbox{.99\linewidth}{%
        Despite the limitations of our research, we present the first measurement of the SMS user report ecosystem, shedding light on the services exploited by scammers to perpetrate this fraudulent activity and de-conflating noisy data such as spam.
    }%
}

\section{Related Work}
\label{sec:relatedwork}

Our work fits into the broader literature on suspicious SMS messages. We divide them by data source.
Previous researchers have investigated public \textit{SMS gateways} to understand the SMS ecosystem and identify SMS phishing text messages.
While Reaves et al.~\cite{reaves2016sending, reaves2018characterizing} and
Moreno et al.~\cite{moreno2023your} mention a minority of SMS messages containing malicious URLs, but they report a wide use of URL shorteners that cannot resolve.
Instead, by enriching our dataset with telemetry captured as SMSs are reported, we gain the ability to, for instance, resolve these shorteners and reliably query relevant contextual information. 
Nahapetyan et al. identify over 67.9$k$ SMS phishing messages using identifiers such as phone numbers, email addresses, and One-Time Codes without removing spam~\cite{nahapetyan2024sms}.

Some have collated victim reports~\cite{twitterreports_eCrime, tang:ccs22} or crowdsourced them~\cite{timko2023commercial, chen_creating_2012}. 
Others used intensive techniques like in-depth interviews~\cite{lutforsmishing_interview, lorrie_phishingsoups} and analyzing news articles~\cite{Oestnews_ecrime}.
More relatedly, some researchers investigated SMS messages caught by XDRs and interacted with the scammers behind the messages, albeit for only one type of scam (Hi Mum/Dad)~\cite{agarwal_usenix25}.
Similarly, Agarwal et al. investigated two months of blocked SMS text messages from one UK \mno and manually categorized them into six scam types~\cite{agarwal_imc24}. Our methods build on their insights into attacker strategies, but we see that just focusing on these six scam types offers an incomplete picture of the landscape. 

It is common to mix together spam and scam messages.
A few groups collect spam and smishing (scam) text messages reported by users on Twitter, % (now known as X),
but treat them all as spam~\cite{tang:ccs22, grier:ccs10}. 
Another recent work crowdsources suspicious text messages from users in the US, categorizing them all as scams~\cite{smishtank_timko24}. 
Srinivasan et al. consider the SMS reports users submit to the Federal Trade Commission (FTC) and third parties as spam~\cite{srinivasan2016understanding}.
Contrastingly, we break down this difference which we show to materially change our results. 
%Our work shows the breakdown of user-reported SMS text messages into OTPs, spam and scam and discusses how SMS with OTPs and phone numbers can be identified as scams or benign. 

Previous work has devised machine learning models to detect smishing messages~\cite{hosseinpour2025poster, jain2020novel, jain2019feature, goel2018smishing, jain2018rule, joo2017s, mishra2020smishing, mishra2019content, sonowal2018smidca}, using old spam datasets~\cite{tiago_spam, chen_creating_2012,delany2012sms}. % and even characterize SMS spear phishing~\cite{smsspearphishing_21}. 
We provide a methodological framework to differentiate spam and identify new scam types from user reports that could help researchers enhance these models.

\section{Conclusion}
\label{sec:conclusion}

 It is mandatory in the UK to reimburse fraud victims after they fall victim to a scam and directly send the fraudster money~\cite{appfraudreimburse}.
 Because of this, UK infrastructure operators are increasingly being pressured by banks to reduce fraud. 
 These incentives have played out in our work investigating hundreds of SMS scams reported during a 4 month period in 2024.
 We winnow down over 1$m$ user reports into 213$k$ scam texts using a careful, multi-layered methodology.
 We investigate these scam texts and find that most that get through the \mnosnospace's XDR have no URL. We hypothesize this is from direct pressure and vast industry experience blocking URL-based scams.
 We highlight the emerging trend of call back scams, where scammers entice victims to call them on a phone number.
We encourage those designing filtering products like XDRs for \mnos to more carefully consider conversation scams, since they often lead to fraud, even if not from the start.

\section*{Acknowledgments}
We would like to thank {\em hidden for anonymous submission}.

\section*{Ethics Considerations}
\label{sec:ethics}

Our work has some ethical considerations, we next explain how we mitigate risks. 
We get access to all 7726 reports submitted by users to a major \mnonospace. 
The first set of safeguard measures is taken by the \mno we collaborate with, who redact users report to anonymize the identity of the users who reports the messages and remove personally identifiable information (PII) from text messages. 
For example, we receive the domain name instead of the complete URL, where a URL might contain personally identifiable information (PII) such as phone number. %Additionally, we do not receive any details about the user who reports the messages, including their mobile number. 
Our collaborators also ensure that our sharing agreement abides by the terms of the 7726 service. 

The second set of safeguards is taken by the authors of this paper, who have designed a protocol and an impact assessment to ensure user reports are stored and processed safely. 
Most critically, we receive and process sender IDs without having explicit consent of the actual sender who has been reported. 
This includes a phone number or an alphanumeric shortcode used to send the message or call a user. 
To protect these users, we avoid any attempts to use this information to identify individuals and we do not interact with any of the phone numbers.
Our experiments are designed to work over aggregates, and our RQs aim at addressing gaps that benefit the research community. 
Our goal is to make strides in enhancing the safety of mobile users --- broader societal benefits --- by providing a deeper understanding of the types of scams being reported and the lure strategies employed by scammers; and by identifying the \mnos that they abuse and their current network status. 
%Following the Belmont report \cite{anabo2019_belmont, miracle2016_belmont}, we determine negligible risks to stakeholders involved, and our work has broader societal benefits. We perform data protection impact assessments to minimize risks. 
Our department's research ethics committee evaluated and approved our measures to minimize risks and this study.

% trigger a \newpage just before the given reference
% number - used to balance the columns on the last page
% adjust value as needed - may need to be readjusted if
% the document is modified later
%\IEEEtriggeratref{8}
% The "triggered" command can be changed if desired:
%\IEEEtriggercmd{\enlargethispage{-5in}}

% references section

% can use a bibliography generated by BibTeX as a .bbl file
% BibTeX documentation can be easily obtained at:
% http://mirror.ctan.org/biblio/bibtex/contrib/doc/
% The IEEEtran BibTeX style support page is at:
% http://www.michaelshell.org/tex/ieeetran/bibtex/
%\bibliographystyle{IEEEtran}
% argument is your BibTeX string definitions and bibliography database(s)
%\bibliography{IEEEabrv,../bib/paper}
%
% <OR> manually copy in the resultant .bbl file
% set second argument of \begin to the number of references
% (used to reserve space for the reference number labels box)

\bibliographystyle{IEEEtran}
\bibliography{references}

%\section{Appendix}
\appendix
\label{appendix}

%\section{Open AI Prompts}
\subsection{Open AI Prompt: Message Classification}
\label{appendix:openai_prompt}

You will receive a json object with an `id' and a `message'. The `id' is the id of the message and the `message' is text of a SMS. Based on the instructions below, process the message and return a json object. 
Instructions: 
1. Identify the brand or organization that the message is trying to impersonate in the text. Return empty if none. ("named\_entity" key in the json object. This key should always be returned in the json.)
2. Classify the type of smishing message ("scam\_type" key in the json. This key should always be returned in the json.) The scam\_types can be: 
"Hey mum/dad" - text addressed to mum/mom or a dad and asking to text/call back potentially giving a reason about using a different mobile number. 
"Delivery/Parcel" - text impersonating a parcel/delivery company asking to click on a link, text back or call on a number
"Banking" - text impersonating a bank or a financial institution asking to click on a link, text back or call on a number
"Government" - text impersonating a government organization asking to click on a link, text back or call on a number
"Telecom" - text impersonating a mobile network operator asking to click on a link, text back or call on a number
"Wrong number" - text with a normal greeting or asking about someone or to reply back
"Spam" - illicit marketing message including casino, betting, random draws, etc
"Others" - If it is does not fit as one of the above category 
3. If the "scam\_type" classified is "Others", then identify a category for the text. Return empty otherwise. ("other\_category" key in the json object. This key should always be returned in the json.)
%4. Extract the phone number if provided in the message asking to text or call back on. Return empty if none. ("phone\_number" key in the json object. This key should always be returned in the json.)
%5. Provide if the message asks to call back or text back as a category. Return empty if none. ("call\_text\_back" key in the json object. This key should always be returned in the json.)
4. Every json object should include the "id" of the message being classified.

\subsection{Open AI Prompt: Detecting Scam Lures}
\label{appendix:openai_prompt_lures}

You will receive a json object with an `id' and a `message'. The `id' is the id of the message and the `message' is text of a scam SMS. Based on the instructions below, process the message and return a json object. 
Instructions: 
1. Provide which lure principles apply for each text message ("lure\_principles" key should be a list and always be provided in the json object. If you cannot detect any lure principles, leave the list empty.) 
Lure principles are: 
a) Distraction Principle - providing various reasons to distract the user.
b) Authority Principle - providing trust to the user to not question authority. could be done by making references to legitimate entities.
c) Herd Principle - encouraging a user to not miss out on opportunities by relating to the popularity of a scheme. convincing how others have won things or take the same risk.
d) Dishonesty Principle - inviting users willingly and knowingly to participate in a fraudulent scheme.
e) Kindness Principle - Fraudsters leverage the willingness of people to help others.
f) Need and Greed Principle - leveraging user's greed and offering attractive (monetary) benefits so user would take an action asked in the text.
g) Time/Urgency Principle - putting time pressure on user so they make a rushed decision.
2. Every json object should include the "id" of the message being classified.

\subsection{Other Scam Types Examples}
\label{appendix:other_scam_types}

\begin{displayquote}
An Enforcement Agent has been scheduled to attend your property, call $<$name$>$ NOW ON $<$phone number$>$ to prevent this action. Quote ref $<$reference number$>$

Can you do this EPC? $<$postcode$>$ Requested day/time: Call to confirm Your fee: £40 Accept job here: [URL]

$<$First Name$>$ is due a Kennel Cough Vaccine on 20/4/24. Call us on $<$phone number$>$ to book. Thank you.

Ensure we hold the correct student finance information for your April payment by visiting:[URL]

$<$customer name$>$, Customer ID: $<$id$>$. You should receive a letter from us soon regarding $<$brand name$>$. It is important that you resolve this matter online as soon as possible by visiting [URL] Thanks, $<$brand name$>$.

[URL] You have not yet paid £39.39 due to $<$brand name$>$ INSURANCE SERVICES GR. Got a question about your account? Text us on $<$phone number$>$. Ref JCF000052164

%We've sent you an urgent email about your student loan. Please read it now. If you don't take action, you could be breaching your Ts and Cs.
\end{displayquote}

\subsection{The Gunning Fog Index}
\label{appendix:gunning_fog_index}

\begin{table}[!ht]
    \centering
    \begin{tabular}{ll}
    \hline
    \multicolumn{1}{c}{\bf Fog Index} & \multicolumn{1}{c}{\bf Reading level by grade}\\
    \hline
    17	& College graduate\\
    16	& College senior\\
    15	& College junior\\
    14	& College sophomore\\
    13	& College freshman\\ 
    12	& High school senior\\
    11	& High school junior\\
    10	& High school sophomore\\
     9	& High school freshman\\
     8	& Eighth grade\\
     7	& Seventh grade\\
     6	& Sixth grade\\
     5	& Fifth grade\\
\hline
\end{tabular}
    \caption{Education level required as per the Gunning fog index.}
    \label{tab:gunning_fog_index}
\end{table}

\begin{figure}[!ht]
    \centering
    \includegraphics[width=0.89\linewidth]{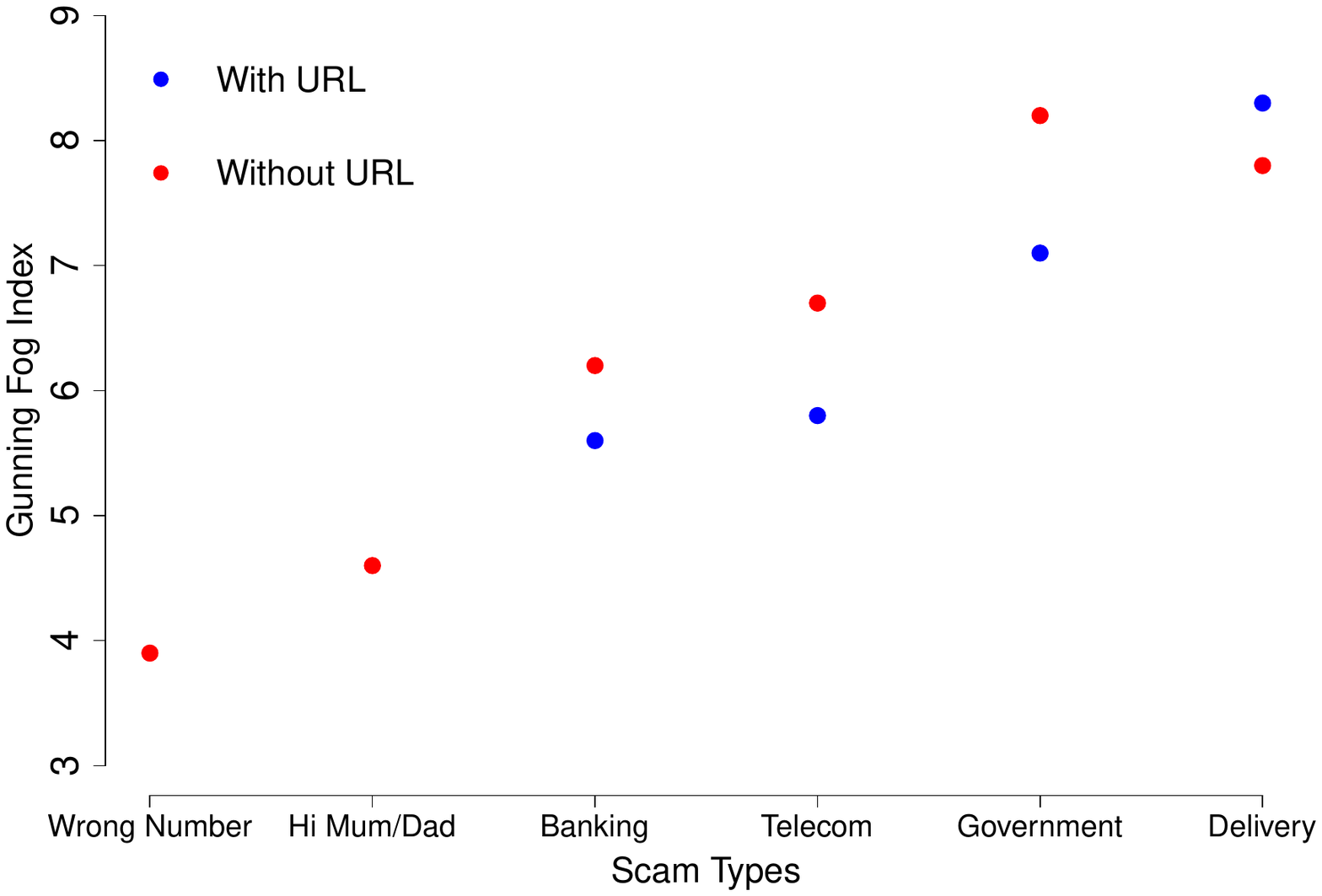}
    \caption{Gunning Fog Index for six scam types with and without URLs.}
    \label{fig:gfi_plot}
\end{figure}

% \subsection{Common Sender Ids for Scam Types}
% \label{appendix:common_sender_ids}

\begin{table*}[!ht]
    \centering
    %\resizebox{\columnwidth}{!}{%
\begin{tabular}{l|rrrrrr}
    \hline
    \multicolumn{1}{c|}{\bf Scam Type} & \multicolumn{1}{c}{\bf Banking} & \multicolumn{1}{c}{\bf Government} & \multicolumn{1}{c}{\bf Delivery/Parcel} & \multicolumn{1}{c}{\bf Telecom} & \multicolumn{1}{c}{\bf Wrong Number} & \multicolumn{1}{c}{\bf Hi Mum/Dad}\\
    \hline
    {\bf Banking} & 6,327\\
    {\bf Government} & 32 & 1,222\\
    {\bf Delivery/Parcel} & 183 & 52 & 11,648 \\
    {\bf Telecom} & 47 & 88 & 57 & 5,742\\
   {\bf Wrong Number} & 47 & 13 & 29 & 35 & 18,404\\
    {\bf Hi Mum/Dad} & 51 & 0 & 52 & 10 & 490 & 8,748\\
    \hline
\end{tabular}
%}
    \caption{Common Sender IDs scammers abuse to send multiple scams.}
    \label{tab:common_senderids}
\end{table*}

% that's all folks
\end{document}